\documentclass[preprint,aps,prc,showpacs,nofootinbib,secnumarabic,superscriptaddress]{revtex4-1}

\usepackage{graphicx}% Include figure files
\usepackage{bm}% bold math
\usepackage{color}
\usepackage{gensymb}
\newcommand{\be}{\begin{equation}}
\newcommand{\ee}{\end{equation}}
\newcommand{\bea}{\begin{eqnarray}}
\newcommand{\eea}{\end{eqnarray}}

\newcommand{\bvec}[1]{\ensuremath{\boldsymbol{#1}}}

\begin{document}

\title{Coherent Hypernucleus Production in Antiproton-Nucleus Annihilation Reactions as a Probe for $\kappa$ meson Exchange}

\author{A.B. Larionov}
\affiliation{Institut f\"ur Theoretische Physik, Universit\"at Giessen,
             D-35392 Giessen, Germany}
\affiliation{National Research Center "Kurchatov Institute", 123182 Moscow, Russia}
\author{H. Lenske}
\affiliation{Institut f\"ur Theoretische Physik, Universit\"at Giessen,
             D-35392 Giessen, Germany}

\date{\today}

\begin{abstract}
  The hypernucleus production reaction $\bar p + {}^AZ \to {}^A_\Lambda(Z-1) + \bar{\Lambda}$ in the beam momentum range $1.5\div20$~GeV/c
  is addressed theoretically as a coherent process.
  The calculations are based on a covariant $t$-channel meson exchange model for the elementary $\bar p p \to \bar\Lambda \Lambda$ annihilation amplitude
  with parameters fixed by comparison with empirical data.
  Besides pseudo-scalar $K$ and vector $K^*$ mesons we also account for correlated $\pi K$ contributions,
  modelled by the scalar $K^*_0(800)$ or $\kappa$ meson.
  Initial and final state nuclear interactions are taken into account in eikonal approximation.
  The bound baryon wave functions are obtained self-consistently in a covariant mean-field approach.
  It is shown that the hypernucleus production cross sections populating discrete states are dominated by the vector and scalar interaction channels.
  The pronounced sensitivity on the scalar $\kappa$ meson exchange contributions indicates that these reactions are well suited
  as a probe for correlated $\pi K$ exchange in the scalar $\kappa/K^*_0$ interaction channel.
\end{abstract}


\maketitle

\section{Introduction}
\label{intro}

Hypernuclei are being produced in manifold ways, by photon-, pion-, antikaon-, proton-, antiproton- and nucleus-nucleus interactions \cite{Bando:1990yi}.
All these types of reactions, except the $\bar p A$ one, are rather well studied both experimentally and theoretically.
But up to now, there are only few theoretical studies of hypernuclear production in antiproton-nucleus interactions
\cite{Bando:1989nx,Cugnon:1990aw,Gaitanos:2011fy,Gaitanos:2014bla,Feng:2015jma,Gaitanos:2016glr}, and all of them address the incoherent production mechanism.
Incoherent hypernucleus production in central collisions is initiated by production of an antikaon in the in-medium $\bar pN$ annihilation
followed by the strangeness exchange process of the type $\bar K N \to Y\pi$. Such reactions are the ideal tool to investigate simultaneously
the production of single- and multi-strangeness systems, as discussed e.g. in the cited works. However, a different approach is required
if spectroscopic studies of the final hypernuclei are the aim. For that purpose, the proper method are peripheral reactions by which hypernuclei
in bound discrete quantum states are obtained. A comprehensive overview of the status of such studies can be found in the recent review of ref. \cite{Gal:2016boi}.
The coherent hypernuclear production in proton- and pion-induced reactions has been investigated in refs. \cite{Shyam:2005mq} and \cite{Bender:2009cj},
respectively, and in photo-induced reactions in ref. \cite{Shyam:2007fm}. Since such coherent reactions are of a perturbative character, they can be described
quantum mechanically by distorted wave methods.

\begin{figure}
\begin{center}
\includegraphics[scale = 0.6]{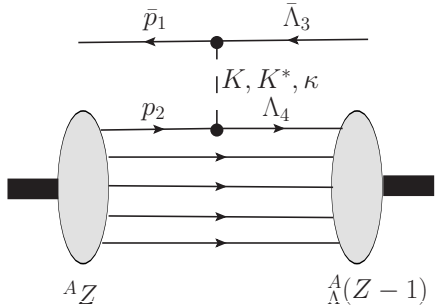}
\caption{\label{fig:hyperProd} The Feynman graph of the ${}^AZ(\bar p,\bar\Lambda){}^A_\Lambda(Z-1)$ process.
  Dashed line represents the propagator of the exchange meson.
  The gray ellipsoids correspond to the wave functions
  of the initial ground state nucleus $^AZ$ and final hypernucleus \mbox{$^A_\Lambda(Z-1)$}.}
\end{center}
\end{figure}

In this paper, we consider $\bar p + {}^AZ \to {}^A_\Lambda(Z-1) + \bar{\Lambda}$ annihilation reactions on a target $^AZ$
leading to the production of a particle-stable hypernucleus $^A_\Lambda(Z-1)$.
The full process is sketched in Fig.~\ref{fig:hyperProd}.
Such reactions could be realized in the foreseeable future at the upcoming \={P}ANDA experiment at FAIR.
Our interest is twofold, namely, first, on the reaction mechanism and, second, on the production dynamics.
The strong coupling of antibaryons to the various annihilation channels requires to account properly for initial (ISI) and final (FSI) state interactions.
For that part we take advantage of our previous study of $\bar p A$ elastic scattering in ref. \cite{Larionov:2016xeb}.
The production proceeds through the elementary $\bar p + p \to \bar \Lambda +\Lambda$ vertex. Since the initial proton and the final $\Lambda$
are constrained to be bound to a nucleus we need to account for the binding potentials. They are described in a relativistic mean-field approach with scalar
and vector fields, similar to the descriptions of hypernuclear bound states in refs. \cite{Bender:2009cj,Glendening:1992du,Keil:1999hk}.

Not much is known, in fact, on the basic $\bar p + p \to \bar \Lambda +\Lambda$ reaction amplitude. Here, we use a meson exchange model.
On the antibaryon side a $\bar u$-quark must be changed into a $\bar s$ quark while on the baryon side a $u$-quark has to be transformed into a $s$-quark.
That can be viewed as the propagation of positively charged mesons of a $[u \bar s]$ quark structure with strangeness $S=1$ from baryon to antibaryon
(or of $[\bar u s]$ mesons with strangeness $S=-1$ in the opposite direction).
Obvious candidates for such a process are the pseudo-scalar ($0^-$) kaon $K$ and vector ($1^-$) $K^*$ mesons.
However, we have to expect that also the $\pi K$ correlated exchange in the scalar $0^+$-channel may play an important role.
The $0^+,S=1$-channel is represented by the $K^*_0(800)$ or $\kappa$ mesons \cite{Agashe:2014kda} which may be considered as the $S=1$ members
of the (hypothetical) scalar meson octet to which also the $\sigma/f_0(600)$ and the $\delta/a_0(980)$ mesons belong.
Like other $0^+$-mesons, the $\kappa/K^*_0$ meson is characterized by a rather broad spectral distribution with uncertain mass and width.
In the present context, the $\kappa$ meson contributes of course through $t$-channel exchange processes. In that sense, we consider the $\kappa$ exchange
as an economical way to take into account the correlated $\pi K$ channel. The $\kappa$ exchange channel is of particular interest for multi-strangeness
baryonic matter in heavy ion collisions and in neutron stars. The $\kappa$ exchange is also an indispensable part of baryon-baryon interaction approaches
utilizing the $SU(3)$-flavour group structure. The Nijmegen group was probably the first one to introduce that channel explicitly \cite{Timmermans:1988hm,Timmermans:1992fu}
into their treatment of baryon-baryon scattering while in the Juelich model that channel is treated dynamically as a $\pi K$-correlation \cite{Haidenbauer:2005zh}.
We note that in the present context the unnatural parity $K$-exchange is strongly suppressed for transition involving bound proton and $\Lambda$ states,
because it is a purely relativistic effect proceeding through the lower wave function components of the Dirac spinors.
Thus, coherent hypernucleus production reactions are perfect tools to addressing specifically the exchange of the natural parity $K^*$ and $\kappa$ mesons.
Any other independent source of information allowing to probe hyperon-nucleon and hyperon-nucleus interactions is highly wanted.
In this respect, hypernuclear reaction physics may provide important clues.

The paper is structured as follows. In section \ref{model} we introduce the Lagrangians describing our covariant annihilation model
and the relativistic mean-field approach for bound baryon states. Both the elementary $\bar p p \to \bar\Lambda \Lambda$
as well as the hypernucleus production $\bar p + {}^AZ \to {}^A_\Lambda(Z-1) + \bar{\Lambda}$ amplitudes
are studied without and with scalar meson exchange. ISI and FSI of the antibaryons in the nucleus
are taken into account in the eikonal approximation. In section \ref{results} the theoretical approach is applied to reactions on an
${}^{40}\mbox{Ar}$ target populating discrete bound states in the $^{40}_{~\Lambda}\mbox{Cl}$ hypernucleus.
Angular distributions and total hypernucleus production cross sections are discussed.
Special attention is paid to the effects introduced by the scalar interaction channel.
In section \ref{SumConcl} we summarize our results and present our conclusions.

\section{The Model}
\label{model}

\subsection{Strangeness production in Antiproton Annihilation Reactions}
The theoretical models for the process $\bar p p \to \bar\Lambda \Lambda$ are divided into two groups: the $t$-channel
strange meson exchange models \cite{Tabakin:1985yv,Kohno:1986mg,LaFrance:1988if,Timmermans:1988hm,Timmermans:1992fu,Haidenbauer:1991kt}
and the quark-gluon models \cite{Rubinstein:1985bb,Furui:1987cc,Burkardt:1988pk,Alberg:1988qu}.
The quark-gluon models are based on the one-gluon ($^3S_1$) or vacuum-type ($^3P_0$) $\bar u u \to \bar s s$ transitions.
Of course, generally, the amplitude of the process $\bar p p \to \bar\Lambda \Lambda$ may be a superposition of the $t$-channel
meson exchanges and the pure quark-gluon transitions. Moreover, based on the existing data currently there is no clear preference
of one type of models over another one.
Thus, we will use here a relatively simple, although well established, $t$-channel meson-exchange framework.

We will introduce the $K$, $K^*$ and $\kappa$ exchanges by using the following
interaction Lagrangians \cite{Cheoun:1996kn,Tsushima:1998jz,Han:1999ck}:
\begin{eqnarray}
   {\cal L}_{KN\Lambda}   &=& -ig_{KN\Lambda} \bar N \gamma^5 \Lambda K + \mbox{h.c.}~,     \label{Lag_KNL}\\
   {\cal L}_{K^*N\Lambda}  &=& \bar N (G_v\gamma^\mu - \frac{G_t}{m_N+m_\Lambda}\sigma^{\mu\nu}\partial_\nu^{K^*})\Lambda K^*_\mu
                                                                  + \mbox{h.c.}~,      \label{Lag_KsNL}\\
   {\cal L}_{\kappa N\Lambda} &=& -g_{\kappa N\Lambda} \bar N \Lambda \kappa  + \mbox{h.c.}~.  \label{Lag_kappaNL}
\end{eqnarray}
The invariant matrix elements for the process $\bar p p \to \bar\Lambda \Lambda$ with the plane wave incoming and outgoing
states can be evaluated by applying standard Feynman rules:
\begin{eqnarray}
  iM_K    &=& -g^2_{KN\Lambda} F^2_K(q^2) \sqrt{\Omega}\,\bar u_{-p_1,-\lambda_1} \gamma^5 u_{-p_3,-\lambda_3}
              \frac{i}{q^2-m_K^2} \bar u_{p_4\lambda_4} \gamma^5 u_{p_2\lambda_2}~,     \label{M_K}\\
  iM_{K^*} &=& -F^2_{K^*}(q^2) \sqrt{\Omega}\,\bar u_{-p_1,-\lambda_1} \Gamma^\mu(-q) u_{-p_3,-\lambda_3}
               iG_{\mu\nu}(q) \bar u_{p_4\lambda_4} \Gamma^\nu(q) u_{p_2\lambda_2}~,     \label{M_Ks}\\
  iM_{\kappa} &=& g^2_{\kappa N\Lambda} F^2_{\kappa}(q^2) \sqrt{\Omega}\,\bar u_{-p_1,-\lambda_1} u_{-p_3,-\lambda_3}
                \frac{i}{q^2-m_{\kappa}^2+im_{\kappa}\Gamma_{\kappa}} \bar u_{p_4\lambda_4} u_{p_2\lambda_2}~, \label{M_kappa}
\end{eqnarray}
where $p_i$ is the four-momentum and $\lambda_i=\pm1/2$ is the spin magnetic quantum number of a
particle $i=1,2,3,4$ (see Fig.~\ref{fig:hyperProd} for the notation),
$q=p_3-p_1$ is the four-momentum transfer. In Eq.(\ref{M_Ks}),
\begin{equation}
   G_{\mu\nu}(q) = \frac{-g_{\mu\nu} + q_\mu q_\nu/m_{K^*}^2}{q^2-m_{K^*}^2+im_{K^*}\Gamma_{K^*}}     \label{G_mu_nu}
\end{equation}
is the $K^*$ meson propagator. The $K^*N\Lambda$ vertex function is defined as
\begin{equation}
   \Gamma^\mu(q)=iG_v\gamma^\mu + \frac{G_t}{m_N+m_\Lambda} \sigma^{\mu\nu} q_\nu~.    \label{Gamma^mu}
\end{equation}
The vertex form factors are chosen in the monopole form:
\begin{equation}
   F_j(q^2) = \frac{\Lambda_j^2-m_j^2}{\Lambda_j^2-q^2}~,~~~j=K,K^*,\kappa~.       \label{FFs}
\end{equation}
Similar to refs. \cite{Sopkovich,Shyam:2014dia,Shyam:2015hqa} we included in Eqs.(\ref{M_K})-(\ref{M_kappa}) the factor $\sqrt{\Omega}$
to describe ISI and FSI where absorption of the flux into other annihilation channels is especially
important. For simplicity, we assume the attenuation factor $\Omega$ to be energy independent.
With $\Omega=1$, Eqs.(\ref{M_K})-(\ref{M_kappa}) correspond to the Born approximation.
The Dirac spinors are normalized according to ref. \cite{BLP}:
$\bar u_{p \lambda} u_{p \lambda} = 2m_{N(\Lambda)}$, $\bar u_{-p,-\lambda} u_{-p,-\lambda} = -2m_{N(\Lambda)}$.

The angular differential cross section in the center-of-mass (c.m.) frame is given by the standard expression:
\begin{equation}
  \frac{d\sigma_{\bar p p \to \bar \Lambda \Lambda}}{d\Omega}
  = \frac{p_{\bar \Lambda \Lambda}}{256\pi^2sp_{\bar p p}}\,
    \sum_{\lambda_1,\lambda_2,\lambda_3,\lambda_4} |M_K+M_{K^*}+M_{\kappa}|^2~,   \label{dsigmadOmega}
\end{equation}
where $s=(p_1+p_2)^2$ is the c.m. energy squared,
$p_{\bar p p}=(s/4-m_p^2)^{1/2}$ and $p_{\bar \Lambda \Lambda}=(s/4-m_\Lambda^2)^{1/2}$ are the c.m. momenta of the
initial and final particles, respectively. 
Note that the interference terms of the kaon exchange amplitude with the
$K^*$ and $\kappa$ exchange amplitudes are equal to zero after summation over spin states
since, in the Born approximation, the unnatural and natural parity exchange amplitudes
do not interfere for unpolarized beam and target (cf. \cite{Tabakin:1985yv}).

The choice of coupling constants is based on $SU(3)$ relations \cite{deSwart:1963pdg}:
\begin{eqnarray}
   g_{KN\Lambda}      &=& -g_{\pi NN} \frac{3-2\alpha_{PS}}{\sqrt{3}}~,      \label{g_KNL}\\
   G_{v,t}           &=& -G_{v,t}^\rho \frac{3-2\alpha_{E,M}}{\sqrt{3}}~,    \label{G_vt}\\
   g_{\kappa N\Lambda}  &=& -g_{\sigma NN} \frac{3-2\alpha_S}{3-4\alpha_S}~,    \label{g_kappaNL}
\end{eqnarray}
where $\alpha$'s are the $D$-type coupling ratios.
The $\pi NN$ coupling constant is very well known, $g_{\pi NN}=13.4$  \cite{Dumbrajs:1983jd}.
The vector $\rho NN$ coupling constant is also fixed, $G_{v}^\rho=2.66$, however, the tensor $\rho NN$ coupling constant
is quite uncertain, $G_{t}^\rho=10.9\div20.6$ \cite{Cheoun:1996kn}.
The $\sigma NN$ coupling constant can be estimated either
from the Bonn model \cite{Machleidt:1987hj} or from the Walecka-type models (cf. \cite{Lalazissis:1996rd}).
In both cases one obtains $g_{\sigma NN} \simeq 10$.
The $\alpha$'s for the octets of light pseudoscalar and vector mesons
are reasonably well determined \cite{Cheoun:1996kn,Han:1999ck}:
$\alpha_{PS} \simeq 0.6$, $\alpha_{E} \simeq 0$, $\alpha_{M} \simeq 3/4$. However, there is no any
phenomenological information on $\alpha_S$.

Thus, the coupling constants $G_{t}$ and $g_{\kappa N\Lambda}$,
the cutoff parameters $\Lambda_K, \Lambda_{K^*}$ and $\Lambda_\kappa$, and the attenuation factor $\Omega$
remain to be determined from comparison with experimental data. We adjusted these parameters to describe
the beam momentum dependence of the total $\bar p p \to \bar\Lambda \Lambda$ cross section. The two sets
of parameters, (1) without $\kappa$ meson and (2) with $\kappa$ meson, are listed in Table~\ref{tab:par}.
In the calculations we used the mass $m_\kappa=682$ MeV and the width $\Gamma_\kappa=547$ MeV \cite{Olive:2016xmw}.
\begin{table}[htb]
  \caption{\label{tab:par}
    Parameters of the $\bar p p \to \bar\Lambda \Lambda$ amplitude.
    The value of $g_{KN\Lambda}$ slightly differs from
    -13.3 as given by Eq.(\ref{g_KNL}) and is taken from
    from $K^+N$ scattering analysis of ref. \cite{Buettgen:1990yw}.
    The cutoff parameters $\Lambda_K$, $\Lambda_{K^*}$ and $\Lambda_\kappa$ are in GeV. In the last column, the attenuation factors are shown.}
  \begin{center}
    \begin{tabular}{|c|c|c|c|c|c|c|c|c|}
    \hline
    Set~~~& $g_{KN\Lambda}$~~~~~& $G_{v}$~~~~~& $G_{t}$~~~~~& $g_{\kappa N\Lambda}$~~~~~& $\Lambda_K$~~~~~& $\Lambda_{K^*}$~~~~~&\
    $\Lambda_\kappa$~~~~~& $\Omega$~~~~~\\
    \hline
    1  & -13.981       &  -4.6     & -8.5     &  ---               &   2.0         &  1.6            & ---             & 0.015 \\
    2  & -13.981       &  -4.6     & -9.0     &  -7.5              &   1.8         &  2.0            & 1.8             & 0.005 \\
    \hline
  \end{tabular}
  \end{center}
\end{table}

\begin{figure}
\begin{center}
   \includegraphics[scale = 0.4]{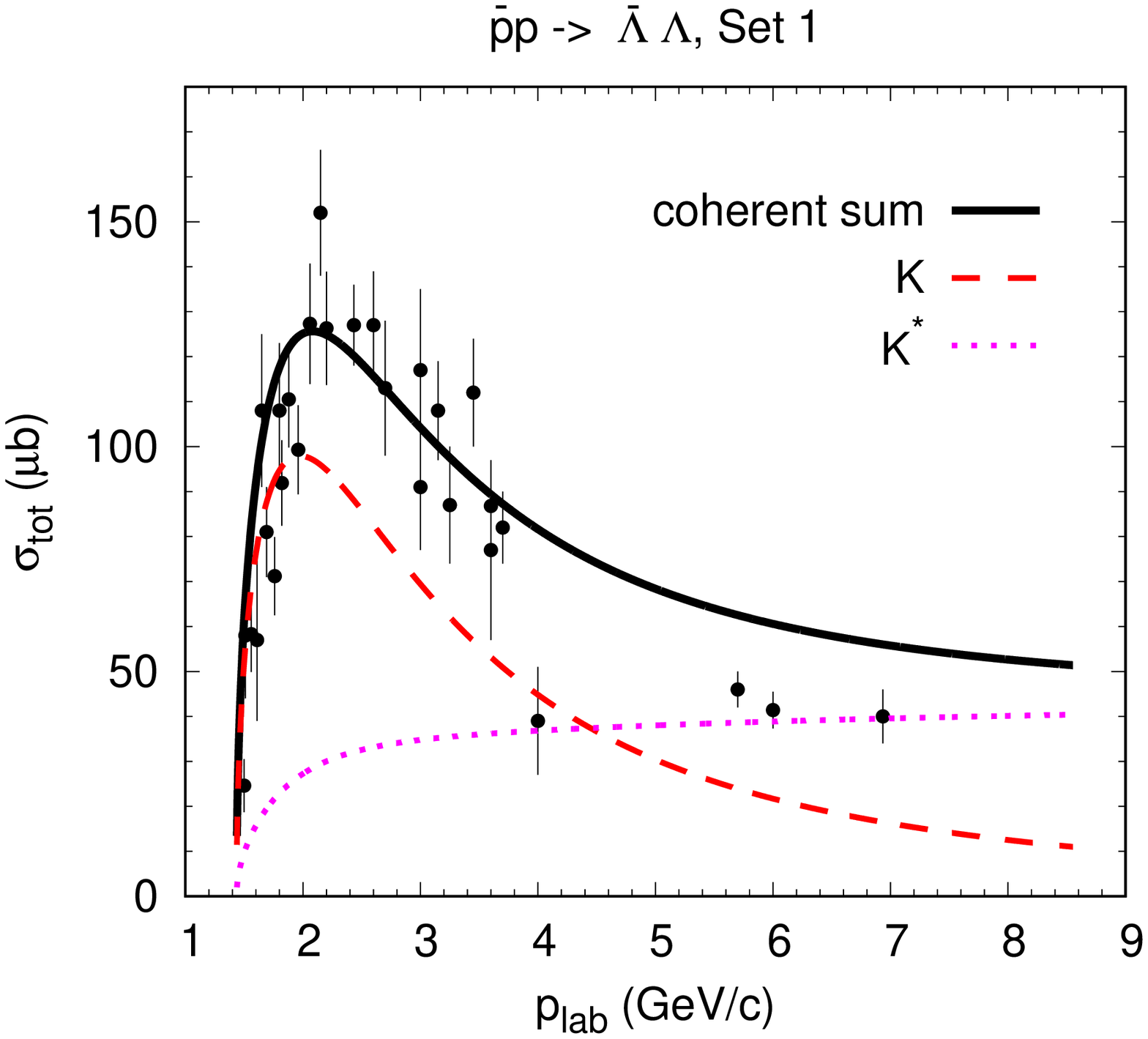}
   \includegraphics[scale = 0.4]{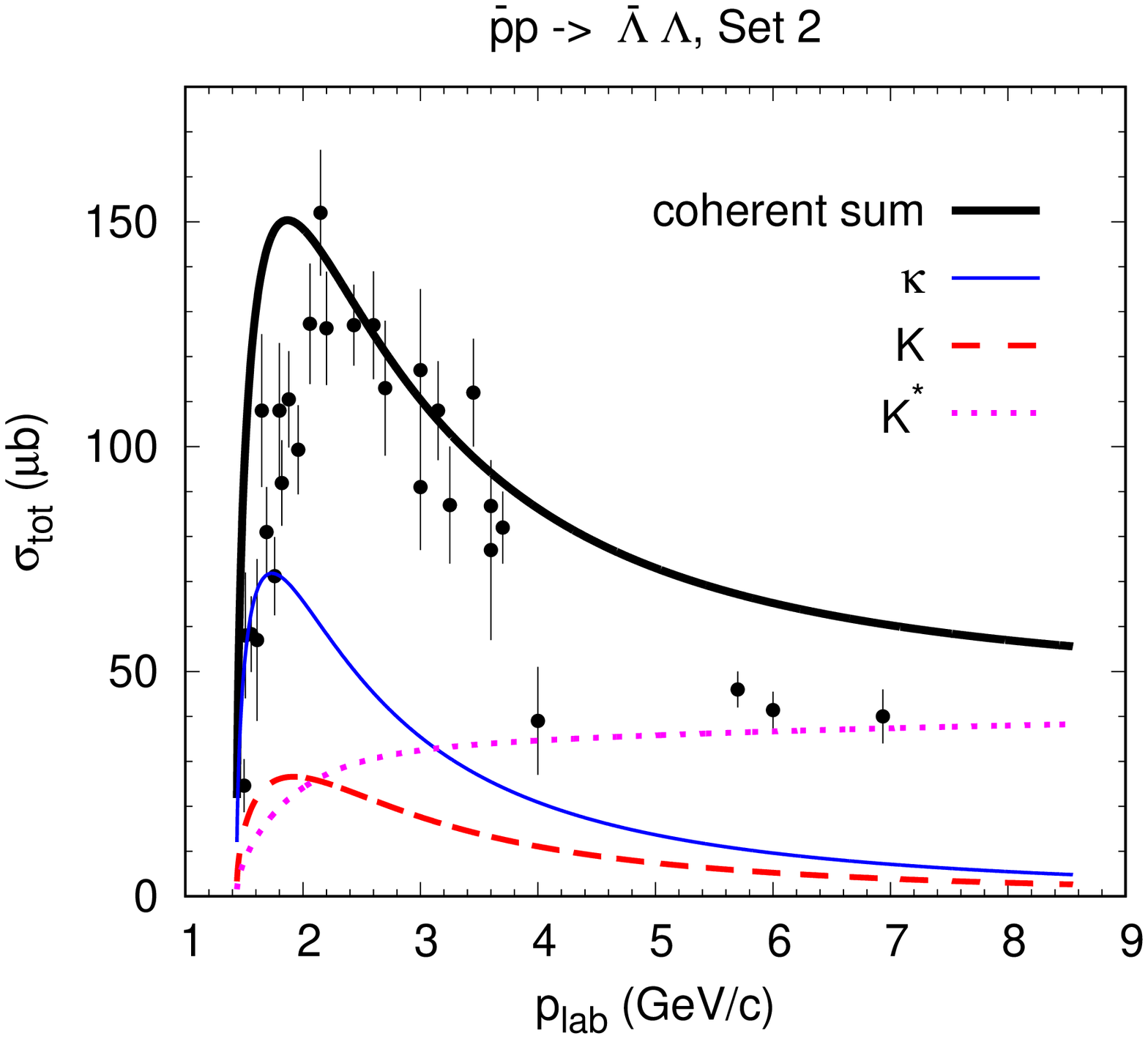}
\end{center}
\caption{\label{fig:sigma_Lbar_L} Total cross section of the process $\bar p p \to \bar\Lambda \Lambda$
  as a function of the beam momentum calculated without (Set 1) and with (Set 2) inclusion of the $\kappa$ meson.
  Experimental data are from ref. \cite{Bald87}.}
\end{figure}
As we see from Fig.~\ref{fig:sigma_Lbar_L}, in the calculation with set 1 the peak of the total $\bar p p \to \bar\Lambda \Lambda$
cross section at $p_{\rm lab} \simeq 2$ GeV/c is saturated by the $K$ exchange.
In contrast, in the case of set 2 the peak is saturated mostly by the $\kappa$ exchange. The $K^*$ exchange contribution grows monotonically
with beam momentum and becomes dominant at $p_{\rm lab} > 3\div4$ GeV/c.
\begin{figure}
\begin{center}
   \includegraphics[scale = 0.4]{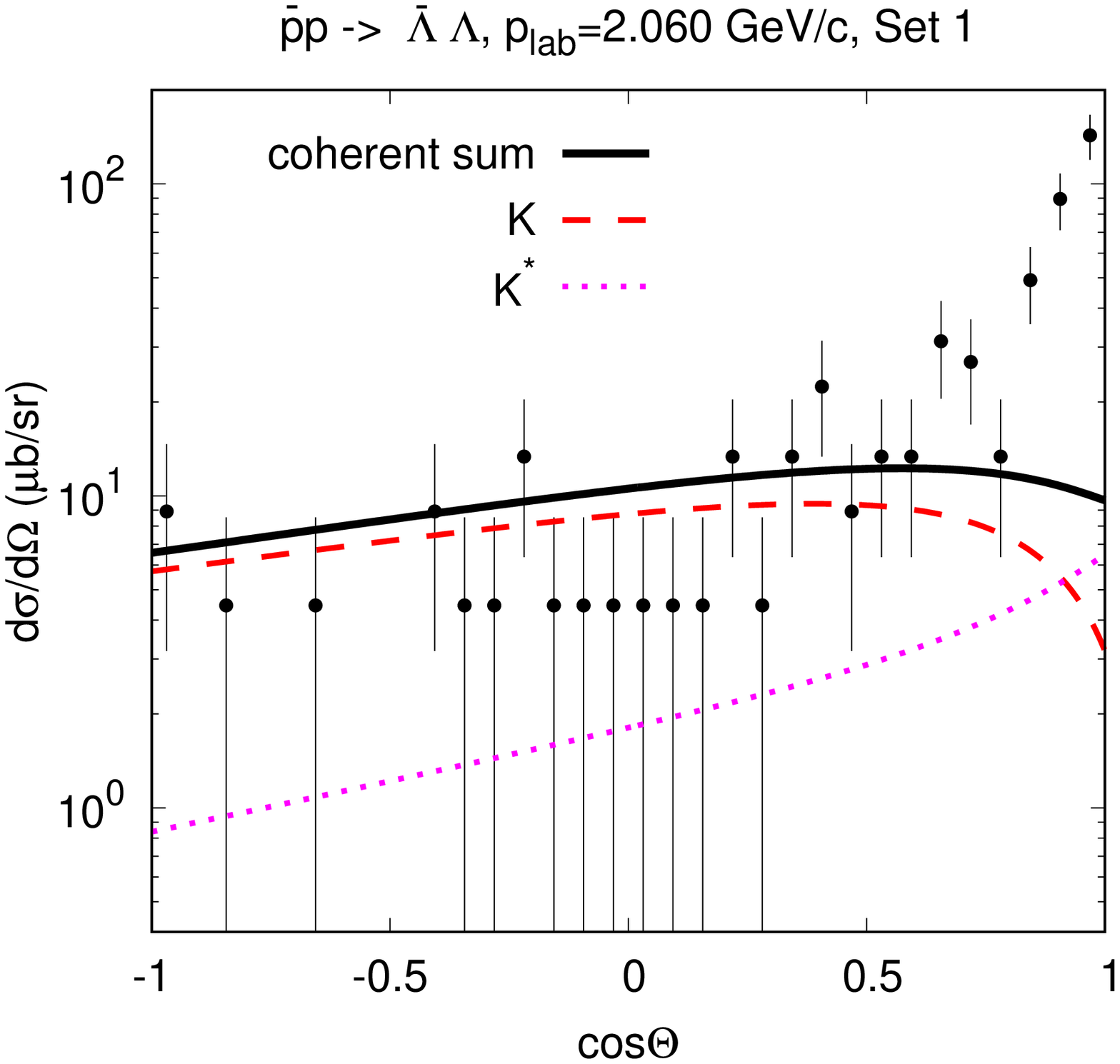}
   \includegraphics[scale = 0.4]{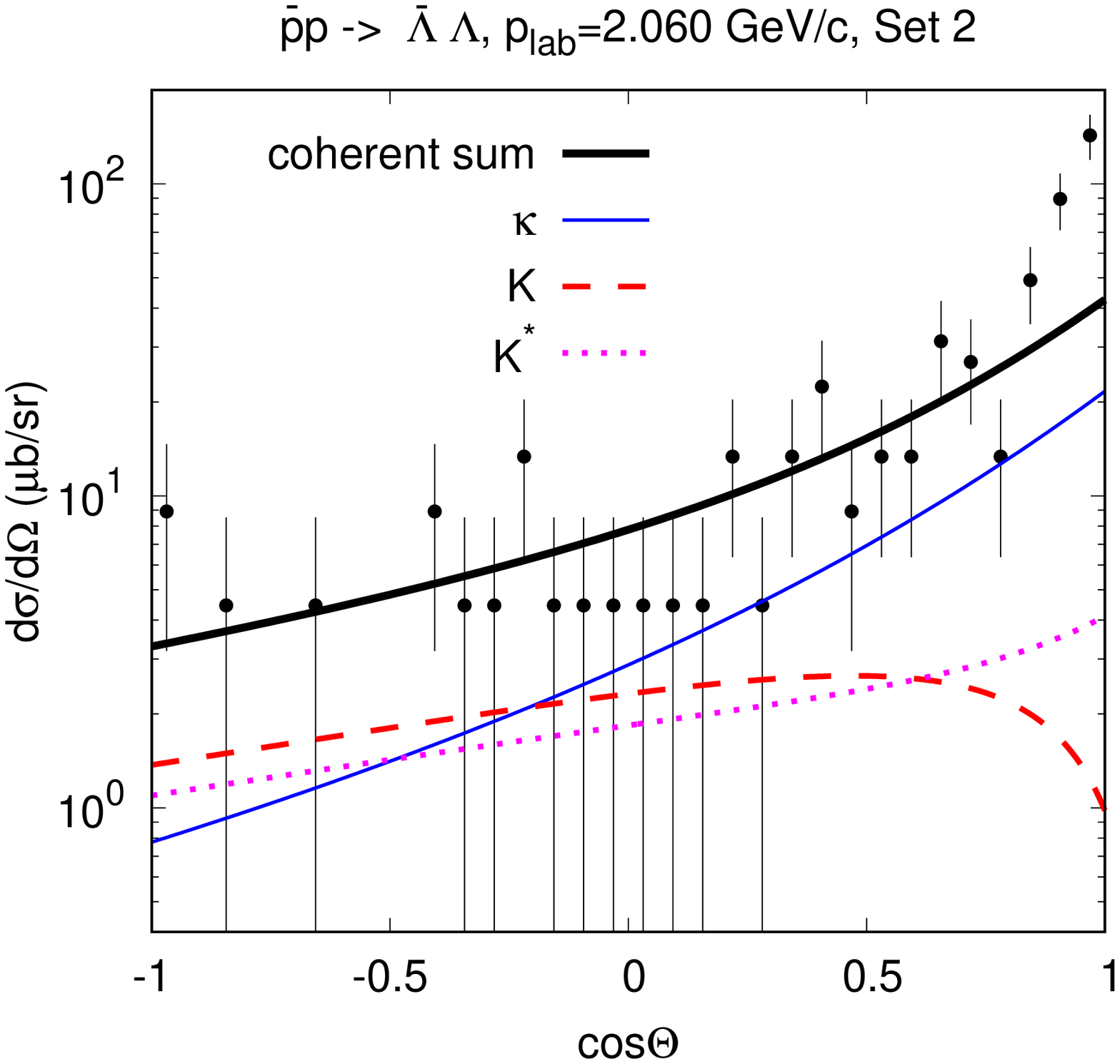}
\end{center}
\caption{\label{fig:dsigdOmega} Angular differential cross section of the process $\bar p p \to \bar\Lambda \Lambda$
  in the c.m. frame at $p_{\rm lab}=2.060$ GeV/c calculated without (Set 1) and with (Set 2) inclusion of the $\kappa$ meson.
  Experimental data are from ref. \cite{Jayet:1978yq}.}
\end{figure}
The effect of the different couplings is better visible in the angular differential cross section displayed in
Fig.~\ref{fig:dsigdOmega}. The kaon exchange contribution to $d\sigma_{\bar p p \to \bar \Lambda \Lambda}/d\Omega$
becomes small at $\Theta=0$ due to the presence of $\gamma^5$ in the matrix element (\ref{M_K}) which interchanges
the upper and lower components of the Dirac spinor\footnote{For the particle at rest the lower component is zero.
Thus, for example, for the elastic $NN$ scattering the parity changing pion exchange contribution vanishes at $\Theta=0$.}.
As a result, at forward c.m. angles the cross section is dominated by $K^*$ and/or $\kappa$ exchange.
Moreover, the latter provides steeper rising differential cross section towards $\Theta=0$ improving
the agreement with experiment.

In the case of the bound proton and $\Lambda$ we include their wave functions in the field operators of the Lagrangians
(\ref{Lag_KNL})-(\ref{Lag_kappaNL}) and calculate the $S$-matrix in the second order perturbation theory using Wick theorem.
After some standard algebra (cf. ref. \cite{BLP}) this leads to the following expression for the $S$-matrix:
\begin{equation}
   S=\frac{2\pi\delta(E_1+E_2-E_3-E_4)}{(2E_1V 2E_3V)^{1/2}} i{\cal M}~,       \label{S}
\end{equation}
where $E_i,~i=1,2,3,4$ are particle energies (see Fig.~\ref{fig:hyperProd} for notation) and $V$ is the normalization volume.
The matrix element ${\cal M}$ in Eq.(\ref{S}) is expressed as a sum of the $K,~K^*$ and $\kappa$ exchange contributions:
\begin{equation}
  {\cal M} = {\cal M}_K + {\cal M}_{K^*} + {\cal M}_{\kappa}~,                  \label{calM}
\end{equation}
where
\begin{eqnarray}
  i{\cal M}_K &=& -g^2_{KN\Lambda} F^2_K(q^2) \sqrt{\Omega}\,\bar u_{-p_1,-\lambda_1} \gamma^5 u_{-p_3,-\lambda_3}
  \frac{i}{q^2-m_K^2}
  \int d^3r e^{-i\bm{q}\bm{r}} \bar \psi_4(\bm{r}) \gamma^5 \psi_2(\bm{r})~,            \label{calM_K}\\
  i{\cal M}_{K^*} &=& -F^2_{K^*}(q^2) \sqrt{\Omega}\,\bar u_{-p_1,-\lambda_1} \Gamma^\mu(-q) u_{-p_3,-\lambda_3}
  iG_{\mu\nu}(q)
  \int d^3r e^{-i\bm{q}\bm{r}} \bar \psi_4(\bm{r}) \Gamma^\nu(q) \psi_2(\bm{r})~,        \label{calM_Ks}\\
  i{\cal M}_{\kappa} &=& g^2_{\kappa N\Lambda} F^2_{\kappa}(q^2) \sqrt{\Omega}\,\bar u_{-p_1,-\lambda_1} u_{-p_3,-\lambda_3}
  \frac{i}{q^2-m_{\kappa}^2+im_{\kappa}\Gamma_{\kappa}}
  \int d^3r e^{-i\bm{q}\bm{r}} \bar \psi_4(\bm{r}) \psi_2(\bm{r})~.                     \label{calM_kappa}
\end{eqnarray}
Here, $\psi_2(\bm{r})$ and $\psi_4(\bm{r})$ are the wave functions of the bound proton and $\Lambda$, respectively.
They satisfy the normalization conditions:
\begin{equation}
   \int d^3r \psi_i^\dag(\bm{r}) \psi_i(\bm{r}) = 1~,~~~i=2,4~.                      \label{normCond}
\end{equation}

The differential cross section in the rest frame of the target nucleus is defined
as follows:
\begin{equation}
  d\sigma = \frac{2\pi\delta^{(4)}(p_1+p_A-p_3-p_B)}{2p_{\rm lab}} \overline{|{\cal M}|^2}
            \frac{d^3p_3}{(2\pi)^32E_3} d^3p_B,                 \label{dSigma}
\end{equation}
where $p_A$ and $p_B$ are the four momenta of the initial nucleus ($A$) and final hypernucleus ($B$).
The $\delta$ function in Eq.(\ref{dSigma}) takes into account the recoil of the hypernucleus.
The averaged modulus squared of the matrix element in Eq.(\ref{dSigma}), i.e. transition probability, is defined as
\begin{equation}
  \overline{|{\cal M}|^2} \equiv \frac{1}{2} \sum_{m,m_\Lambda,\lambda_1,\lambda_3} |{\cal M}|^2~,   \label{calM2}
\end{equation}
where $m$ and $m_\Lambda$ are the spin magnetic quantum numbers of the occupied proton state from the valence
shell and of the $\Lambda$ hyperon, respectively, and the factor of $1/2$ expresses the averaging over $\lambda_1$.

The matrix elements (\ref{calM_K})-(\ref{calM_kappa}) are obtained in the impulse approximation (IA).
More realistic calculation should take into account the distortion of the incoming $\bar p$ and outgoing $\bar\Lambda$
waves, mostly due to strong absorption of the antibaryons in the nucleus. In the eikonal approximation the incoming
$\bar p$ wave is multiplied by the factor
\begin{equation}
  F_{\bar p}(\bm{r}) =
  \exp\left(-\frac{1}{2}\sigma_{\bar pN}(1-i\alpha_{\bar pN}) \int\limits_{-\infty}^0 d\xi
        \rho(\bm{r}+\frac{\bm{p}_{\bar p}}{p_{\bar p}}\xi)\right)~,   \label{F_barp}
\end{equation}
and the outgoing $\bar\Lambda$ wave is multiplied by
\begin{equation}
  F_{\bar\Lambda}(\bm{r}) =
  \exp\left(-\frac{1}{2}\sigma_{\bar \Lambda N}(1-i\alpha_{\bar \Lambda N}) \int\limits_0^{+\infty} d\xi
        \rho(\bm{r}+\frac{\bm{p}_{\bar\Lambda}}{p_{\bar\Lambda}}\xi)\right)~,
                     \label{F_barL}
\end{equation}
where $\rho(\bm{r})$ is the nucleon density, $\sigma_{jN}$ is the total $jN$ cross section,
$\alpha_{jN}=\mbox{Re}f_{jN}(0)/\mbox{Im}f_{jN}(0)$ is the ratio of the real-to-imaginary part of the forward $jN$ amplitude
($j=\bar p, \bar\Lambda$).
Equations (\ref{F_barp}),(\ref{F_barL}) can be obtained by applying the eikonal approximation to solve the Schr\"odinger equation
for the scattering of a particle in the external potential (cf. ref. \cite{LL}) which is then replaced by the optical potential
in the low-density approximation. Since the factors $F_{\bar p}(\bm{r})$, $F_{\bar\Lambda}(\bm{r})$ are weakly changed on the
distances $\sim m_K^{-1}$, the $S$-matrix can be calculated in the local approximation which results in multiplying the integrands
in the matrix elements (\ref{calM_K})-(\ref{calM_kappa}) by $F_{\bar p}(\bm{r}) F_{\bar\Lambda}(\bm{r})$. (Similar expressions can be
also found, e.g., in refs. \cite{Bando:1990yi,Frankfurt:1994nn}.) In numerical calculations we applied the momentum dependent total $\bar pN$ cross
section and the ratio $\alpha_{\bar p N}$ as described in ref. \cite{Larionov:2016xeb}. We have assumed that
$\sigma_{\bar\Lambda N} = \sigma_{\bar p N}$ at the same beam momenta which is supported by experimental data on the total $\bar\Lambda p$
cross section at $p_{\rm lab}=4\div14$ GeV/c \cite{Eisele:1976fe}. For simplicity we have set $\alpha_{\bar \Lambda N}=0$.

Note that the factor $\sqrt{\Omega}$ in Eqs.(\ref{calM_K})-(\ref{calM_kappa}) expresses the modification of the elementary
$\bar p p \to \bar\Lambda \Lambda$ amplitude due to ISI and FSI in the colliding system.
However, the factor of $F_{\bar p}(\bm{r}) F_{\bar\Lambda}(\bm{r})$ takes into account the modification of the
$^AZ(\bar p,\bar\Lambda)\,^A_\Lambda(Z-1)$ amplitude due to sequential elastic rescattering of the incoming $\bar p$ and outgoing
$\bar \Lambda$ on the different nucleons.

\subsection{Nuclear Structure Aspects}
In agreement with the covariant formulation of the production vertices the nucleon and hyperon single particle bound state wave functions are determined
as solutions of a static Dirac equation with scalar and vector potentials, similar to refs. \cite{Glendening:1992du,Keil:1999hk,Bender:2009cj}.
The baryon Dirac-spinors are obtained from the fermion wave equation (cf. \cite{BLP}):
\begin{equation}\label{statDirac}
\left( -i \bvec{\alpha}\cdot \bvec{\nabla} + \beta m^*_B(r) + V_B(r)+q_BV_C(r)- \varepsilon \right) \psi_B(\bm{r}) =0,
\end{equation}
where $m^*_B(r)=m_B+S_B(r)$ is the effective (Dirac) mass.
Both the scalar ($S_B$) and nuclear vector ($V_B$) potentials are in general superpositions of
the classical meson fields $U_{BM}$, weighted by the strong interaction coupling constants appropriate for the given baryon.
Here, $M=\sigma(I=0,J^{P}=0^+),~\omega(0,1^-),~\delta(1,0^+),~\rho(1,1^-)$ stands for the meson mediating the interaction in the respective channel.
For the nucleons the scalar and vector potentials are defined as 
\begin{eqnarray}
  S_N(r) &=& U_{N\sigma}(r) + U_{N\delta}(r) \tau^3~,  \label{S_N}\\
  V_N(r) &=& U_{N\omega}(r) + U_{N\rho}(r) \tau^3~,  \label{V_N}
\end{eqnarray}
where $\tau^3=+1(-1)$ for the neutron (proton).
For charged particles with charge $q_B$ also the static Coulomb potential ($V_C$) contributes \cite{Keil:1999hk}.
The meson fields are parameterized by Woods-Saxon (WS) form factors:
\be
  U_{NM}(r) = \frac{U_{NM}^{(0)}}{e^\frac{r-R_M}{a_M}+1}~.      \label{U_WS}
\ee
Assuming spherically symmetric potentials the eigenfunctions of the Dirac equation are characterized by radial, orbital and total angular momentum quantum numbers,
$n,~l,~j$, respectively, together with the magnetic quantum numbers $m \equiv j_z$. The spinors are given by the upper and lower Pauli-type components
\begin{equation}
  \psi_{nljm}(\bm{r})=
  \left(
  \begin{array}{l}
    f_{nlj}(r) \mathcal{Y}_{jm}^l(\Theta,\phi)\\
    i g_{nlj}(r) \mathcal{Y}_{jm}^{l^\prime}(\Theta,\phi)
  \end{array}
  \right)~,       \label{psi_nljm}
\end{equation}
where $l^\prime=2j-l$, and $\mathcal{Y}_{jm}^l(\Theta,\phi)$ denotes the spherical spin-orbit spinor \cite{VMKh}.

During the calculation, the strength factors $U_{NM}^{(0)}$ and the geometrical parameters $R_M, a_M$
are considered as global variational parameters.
They are determined self-consistently by the constraint to reproducing nuclear binding energies and nuclear root-mean-square radii.
The fitted potentials correspond to the full self-energies, including rearrangement contributions.
Nuclear binding energies are calculated by projecting out the rearrangement self-energy contributions \cite{Lenske:1995wyj}.

In the spirit of the relativistic mean-field (RMF) approach, the volume integrals of the rearrangement-corrected potentials 
are related to the density-averaged meson-baryon coupling constants as follows:
\be
 g^2_{MNN} = (-1)^{J+1} m_M^2\, \frac{\int d^3r U_{NM}(r)}{\int d^3r \rho_M(r)}~,    \label{eq:Coupling} 
\ee
where $m_M$ is the meson mass. The source densities of the meson fields are determined as the expectation values of
the nucleon field, $\psi(\bm{r})$, operator products:
$\rho_\sigma(r)=\langle\bar\psi(\bm{r})\psi(\bm{r})\rangle$,
$\rho_\omega(r)=\langle\psi^\dag(\bm{r})\psi(\bm{r})\rangle$,
$\rho_\delta(r)=\langle\bar\psi(\bm{r}) \tau^3 \psi(\bm{r})\rangle$,
$\rho_\rho(r)=\langle\psi^\dag(\bm{r}) \tau^3 \psi(\bm{r})\rangle$.

The hyperon self-energies are defined correspondingly. In that case the vertex is given by a product of coupling constant $g^2_{MNN} \to g_{MYY}g_{MNN}$.
As in \cite{Keil:1999hk} we define the scaling factor $R_{YM}=g_{MYY}/g_{MNN}$ which allows to write the hyperon potentials in leading order 
as $U_{YM}(r)=R_{YM}U_{NM}(r)$. 
Since the $\Lambda$ hyperon is an uncharged isoscalar particle, its scalar and vector potentials contain only isoscalar components,
i.e. $S_{\Lambda}(r)=U_{\Lambda\sigma}(r)$ and $V_{\Lambda}(r)=U_{\Lambda\omega}(r)$.

\section{Application to Hypernucleus Production on an ${}^{40}\mbox{Ar}$ Target}
\label{results}

As a representative case we consider the reaction
$
   \bar p + {}^{40}\mbox{Ar} \to \bar\Lambda + {}^{40}_{~\Lambda}\mbox{Cl}
$.
The choice of the ${}^{40}\mbox{Ar}$ target is motivated by the future \={P}ANDA experiment at FAIR where noble gases will be
used as targets\footnote{For lighter nuclei, such as ${}^{20}\mbox{Ne}$, the recoil corrections should be taken into account
in more detail, cf. \cite{Bando:1990yi}.}.

The WS parameters of the scalar and nuclear vector potentials in that mass region are displayed in Table~\ref{tab:SelfE} where also the derived coupling constants 
for standard values of the meson masses are shown. It is seen that the self-consistently derived values of the $\sigma NN$ and $\omega NN$ coupling constants
are almost perfectly agreeing with the values used in other RMF approaches, e.g. the widely used NL3-parameter set \cite{Lalazissis:1996rd}.
However, here we include also the otherwise often neglected scalar-isovector interaction channel, represented by the $\delta/a_0(980)$ meson, which is important 
to keep track of the mass evolution far off beta-stability.
Since the signs of the scalar-isovector and vector-isovector fields are opposite, these two fields largely compensate each other. Thus, the $\rho NN$ coupling constant
is larger than that of NL3 (see also the dedicated study of nuclear matter properties in the RMF models with and without $\delta$ meson  in ref. \cite{Liu:2001iz}).
\begin{table}
\begin{center}
\begin{tabular}{|c|c|c|c|c|c|c|}
  \hline
        Self-Energy      & $U_{NM}^{(0)}$ [MeV]&  $r_{0,M}$ [fm] & $a_M$ [fm]& Meson     & Mass [MeV]&  $g^2_{MNN}/4\pi$ \\ \hline
        scalar-isoscalar &             -402.0&  1.0806        & 0.553     &  $\sigma$ &   550     &  8.1179 \\ 
        vector-isoscalar &              328.0&  1.0700        & 0.520     &  $\omega$ &   783     &  12.8052 \\ 
        scalar-isovector &     $-80.0 \alpha$&  1.1800        & 0.500     &  $\delta$ &   980     &  6.3037\\ 
        vector-isovector &     $ 90.0 \alpha$&  1.1500        & 0.520     &  $\rho$   &   775     &  4.1794\\
  \hline
\end{tabular}
  \caption{\label{tab:SelfE} Nucleon mean-field potentials and meson-nucleon coupling constants. 
    The potential radii in Eq.(\ref{U_WS}) are expressed as $R_M=r_{0,M}A^{1/3}$. The coupling constants are defined by Eq.\protect{(\ref{eq:Coupling})}.
    Note that the isovector potentials include the isospin asymmetry factor of the nucleus $\alpha=(N-Z)/A$.}
\end{center}
\end{table}
The calculated binding energy of the ${}^{40}\mbox{Ar}$ nucleus is $B=343.58$~MeV, which compare very well to the value from the AME compilation
\cite{Audi:2014eak}, $B_{exp}=343.81$~MeV. The r.m.s radii of proton and neutron density distributions are, respectively,
$\sqrt{\langle r^2 \rangle_p}=3.30$ (3.33) fm, and $\sqrt{\langle r^2 \rangle_n}=3.41$ (3.43) fm, where the phenomenological values from
the Skyrme-Hartree-Fock systematics are given in brackets. Without going into details we mention that after a very modest, $\sim 0.01\%$,
modification of the radius parameter $r_{0,\sigma}$ the binding energies of the neighboring isotopes, i.e. ${}^{39}\mbox{Cl}$ and ${}^{39}\mbox{Ar}$,
are reproduced, thus describing properly also the proton and neutron separation energies in ${}^{40}\mbox{Ar}$.

Under the assumption that the nuclear potentials do not change after a sudden removal of the valence proton, the $\Lambda$-hyperon scalar and vector potentials
in the ${}^{40}_{~\Lambda}\mbox{Cl}$ nucleus were obtained by multiplying the scalar and vector nucleon potentials in the ${}^{40}\mbox{Ar}$ nucleus by the factors
$R_{\Lambda\sigma}=0.525$ and $R_{\Lambda\omega}=0.550$, respectively. This leads to a good agreement of the $\Lambda$ energy levels with the empirical
systematics and with the previous relativistic mean-field calculations \cite{Keil:1999hk}, as seen from Table~\ref{tab:Lambda_bind}.
\begin{table}[htb]
  \caption{\label{tab:Lambda_bind}
    Binding energies of the $\Lambda$ states in the ${}^{40}_{~\Lambda}\mbox{Cl}$ nucleus.
    Empirical $\Lambda$ binding energies (spin-orbit splitting not resolved) for ${}^{40}_{~\Lambda}\mbox{Ca}$
    from ref. \cite{Keil:1999hk} are given in brackets.}
  \begin{center}
    \begin{tabular}{|l|l|}
    \hline
    $\Lambda$ state~~~& $B_\Lambda$ [MeV]~~~~~\\
    \hline
    $1s_{1/2}$  &  18.55 ($18.7\pm1.1$)  \\
    $1p_{3/2}$  &  10.20 ($9.9\pm1.1$)   \\
    $1p_{1/2}$  &   9.26 ($9.9\pm1.1$)   \\
    $1d_{5/2}$  &   2.14 ($1.5\pm1.1$)   \\
    $2s_{1/2}$  &   1.44  \\
    $1d_{3/2}$  &   0.84 ($1.5\pm1.1$)   \\
    \hline
  \end{tabular}
  \end{center}
\end{table}

In order to assure that after the reaction the residual core nucleus carries as little excitation energy as possible,
we consider only strangeness creation processes on protons of the ${}^{40}\mbox{Ar}$ $1d_{3/2}$ valence shell.

\begin{figure}
\begin{center}
  \includegraphics[scale = 0.4]{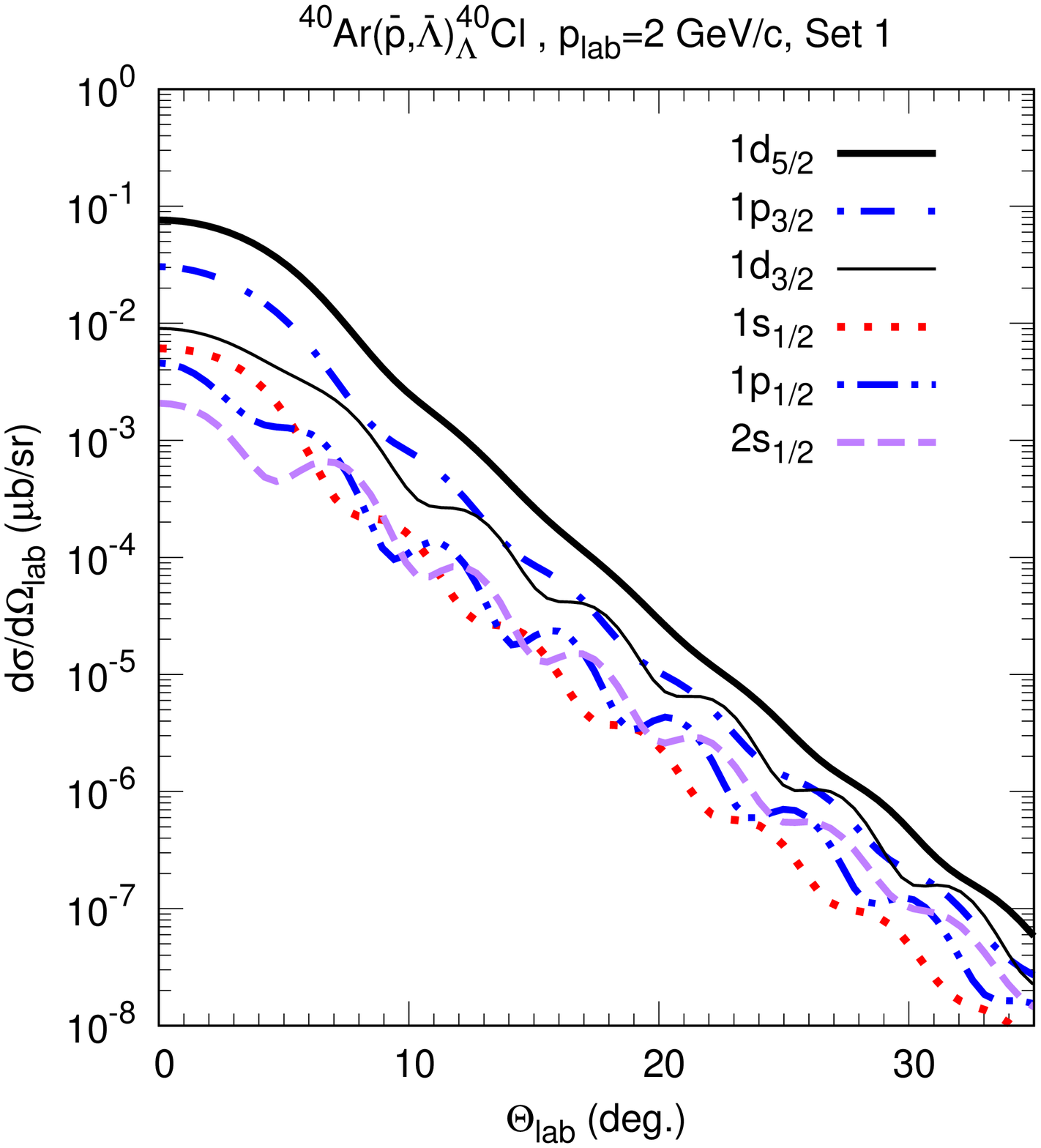}
  \includegraphics[scale = 0.4]{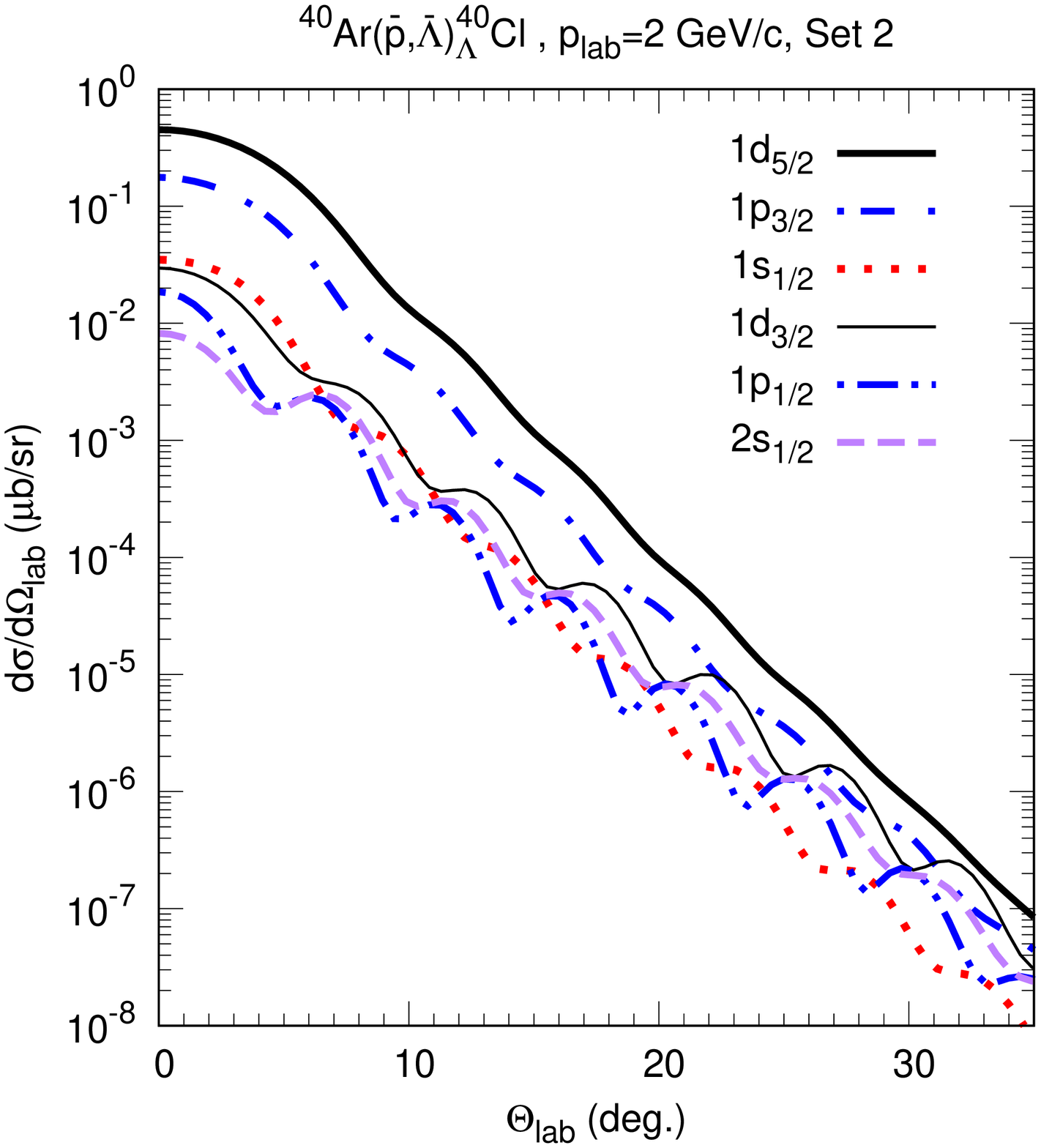}
\end{center}
\caption{\label{fig:dsigdO_sum} Angular differential cross section of the reaction ${}^{40}\mbox{Ar}(\bar p,\bar \Lambda){}^{40}_{~\Lambda}\mbox{Cl}$
  at $p_{\rm lab}=2$ GeV/c. Lines show the calculations for $\Lambda$ in various states, as indicated. Left and right panels display calculations
without (Set 1) and with (Set 2) $\kappa$ exchange.}
\end{figure}

The differential hypernuclear production cross sections with the $\Lambda$ occupying various shells are compared in Fig.~\ref{fig:dsigdO_sum}.
Irrespective of spin-orbit effects, overall the cross sections are larger for larger hyperon orbital angular momentum, i.e. $l_\Lambda$.
This is a consequence of the interplay of several effects:
\begin{itemize}
\item The momentum transfer at $\Theta=0$ is small ($\sim 0.3$ GeV/c) implying a suppression of the $p \to \Lambda$ transitions with large orbital momentum transfer.
\item The number of the spin states of the $\Lambda$ contributing to the transition probability of Eq.(\ref{calM2}), i.e. $2(2l_\Lambda+1)$, grows obviously with $l_\Lambda$.
\item For $\Lambda$ states with larger $l_\Lambda$ the hyperon probability distribution is increasingly shifted to larger radii.
      Hence, the absorption effects are diminished  with increasing $l_\Lambda$.
\end{itemize}
The inclusion of $\kappa$ exchange leads to significant enhancement of the cross sections at small polar angles for all states of the produced hypernucleus
which is also expected from Fig.~\ref{fig:dsigdOmega}.

\begin{figure}
\begin{center}
   \includegraphics[scale = 0.4]{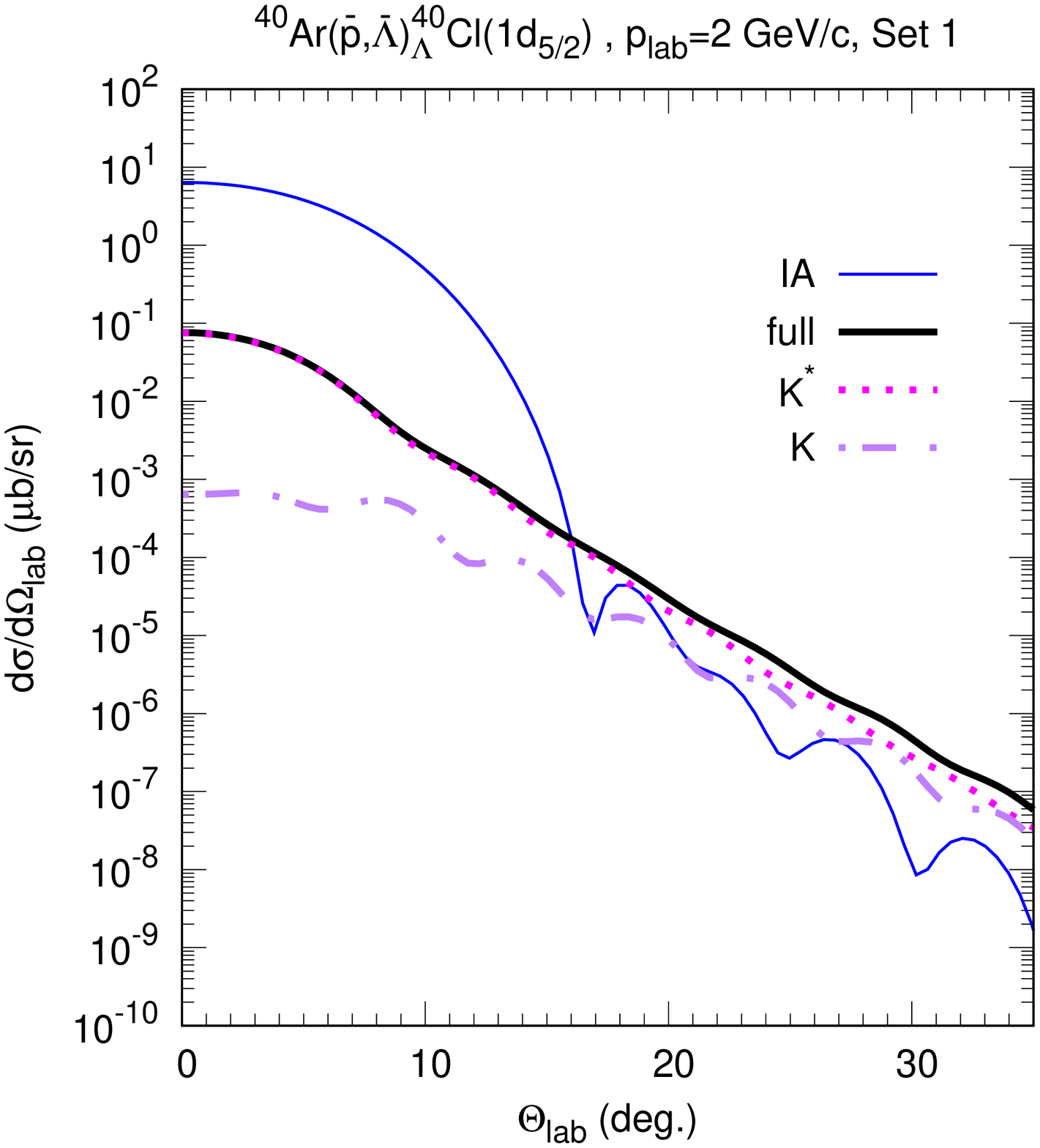}
   \includegraphics[scale = 0.4]{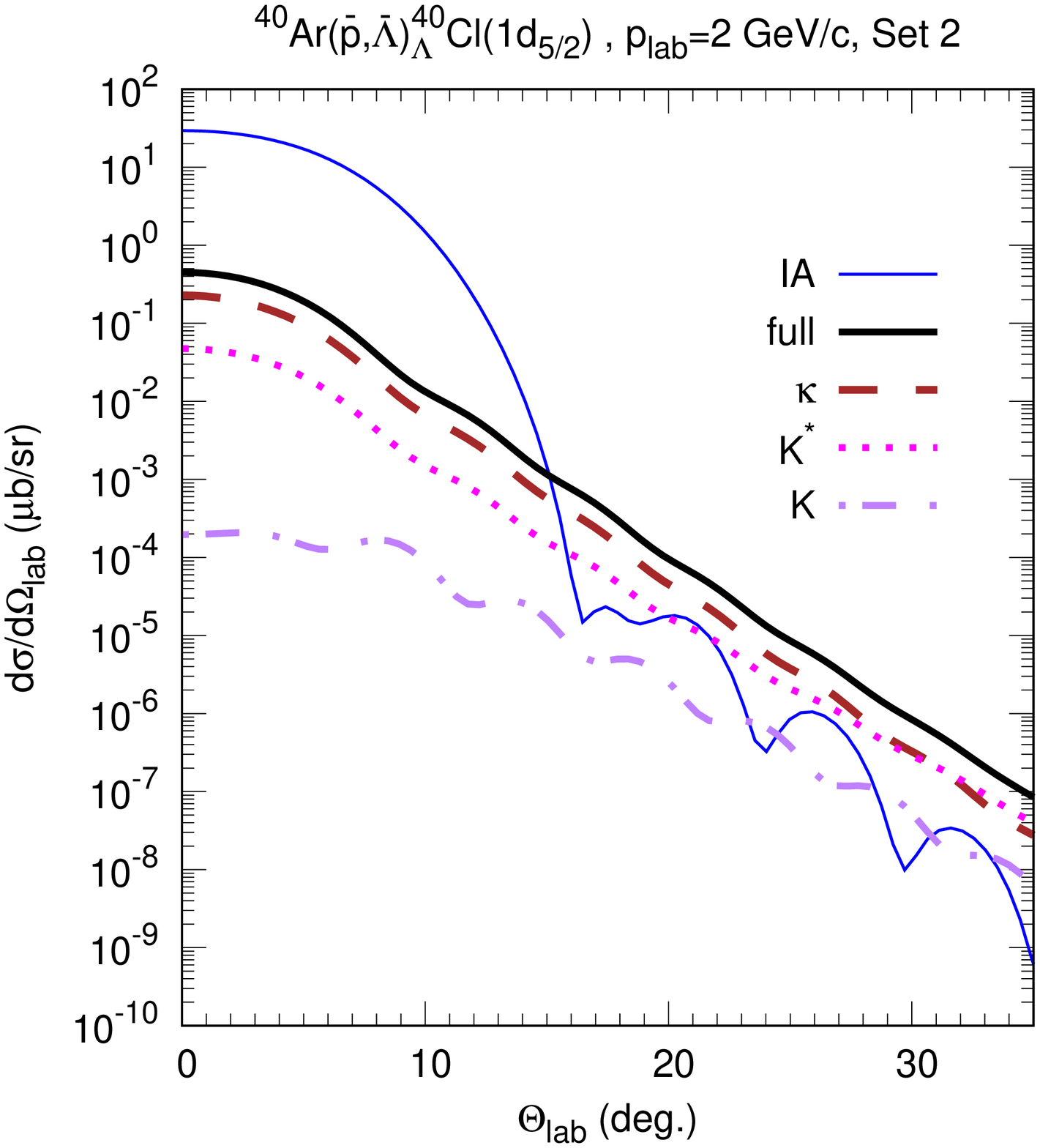}
\end{center}
\caption{\label{fig:dsigdO_1d2.5} Angular differential cross section of the reaction ${}^{40}\mbox{Ar}(\bar p,\bar \Lambda){}^{40}_{~\Lambda}\mbox{Cl}$
  at $p_{\rm lab}=2$ GeV/c with $1d_{5/2}$ $\Lambda$ state.
  As indicated, the IA calculation, the full calculation (with absorption), and the separate meson contributions
  to the full calculation are shown by different lines.
  The left and right panels display the results without (Set 1) and with (Set 2) $\kappa$ meson, respectively.}
\end{figure}
The largest cross section is obtained for the ${}^{40}_{~\Lambda}\mbox{Cl}$ hypernucleus with $\Lambda$ in the $1d_{5/2}$ state.
The differential angular distribution for this case is analyzed in more detail in Fig.~\ref{fig:dsigdO_1d2.5}.
From the comparison of the full and IA calculations we observe that the absorption of $\bar p$ and $\bar \Lambda$ has a quite significant effect:
it reduces the cross section drastically, amounting at forward angles
to about two orders of magnitude, and smears out the diffractive structures.
Similar effects of the absorption are present also for the other $\Lambda$ states (not shown).

A deeper insight into the production mechanism is obtained by decomposing the total reaction amplitude into different meson exchange parts.
From the partial meson exchange contributions, shown in Fig.~\ref{fig:dsigdO_1d2.5}, it is remarkable that for Set 1 the kaon contribution is small
and the spectrum is dominated by $K^*$, even at large angles, while , on first sight, from Figs.~\ref{fig:sigma_Lbar_L},\ref{fig:dsigdOmega}
one would expect the opposite. For example, for $\bar p A$ collisions at $p_{\rm lab}=2$ GeV/c the $\bar \Lambda$ produced at $\Theta_{\rm lab}=30\degree$
carries away the momentum transfer of $\sim 1$ GeV/c . This corresponds approximately to $\Theta=90\degree$ in c.m. frame if translated
into the $\bar p p \to \bar\Lambda \Lambda$ reaction in free space. Thus, we should expect (see left Fig.~\ref{fig:dsigdOmega}) that the kaon exchange
should be a factor of five larger than the $K^*$ exchange.

However, in the case of the nuclear target the $K^*$ exchange contribution is larger than that of $K$ exchange even at $\Theta_{\rm lab}=30\degree$.
This surprising result can be understood by the fact that the momentum transfer to the $\bar\Lambda$ is provided by the nucleus as a whole while
the hyperon is almost at rest. The exchange by pseudoscalar meson is suppressed in this case since it proceeds through the lower components
of the proton and $\Lambda$ Dirac spinors which are suppressed by factors $\sim 1/m_BR$, where $R$ is the nuclear radius.
In contrast, in the case of the free space $\bar p p \to \bar\Lambda \Lambda$ process at $\Theta=90\degree$
the $\Lambda$ is produced with momentum $\sim 1$ GeV/c and, thus, the upper and lower components of its Dirac spinor
are of comparable magnitude which favors the pseudoscalar meson exchange.

The situation is very different in the case of Set 2. Here, $\kappa$ plays the dominant role both for the free scattering
$\bar p p \to \bar\Lambda \Lambda$ and for the hypernucleus production since scalar exchange is not suppressed in recoilless kinematics.

\begin{figure}
\begin{center}
   \includegraphics[scale = 0.4]{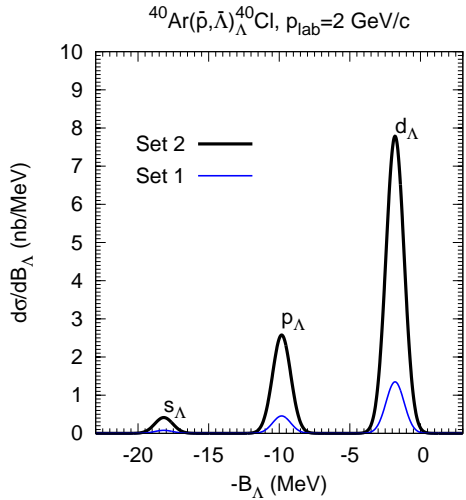}
\end{center}
\caption{\label{fig:Ar40_spectrum} The $\Lambda$ binding energy spectrum of the ${}^{40}_{~\Lambda}\mbox{Cl}$ hypernuclei
  coherently produced in $\bar p\,{}^{40}\mbox{Ar}$ collisions at $p_{\rm lab}=2$ GeV/c.
  The smooth curves are obtained by multiplying the angle-integrated cross sections for the hypernucleus production
  in $1s_{1/2}$, $1p_{3/2}$ and $1d_{5/2}$ states by the Gaussians of a width FWHM=1.5 MeV which is
  a typical experimental energy resolution.}
\end{figure}

As we see from Fig.~\ref{fig:Ar40_spectrum}, the cross section of coherent hypernucleus production in the different states is much larger
when the $\kappa$ exchange is included. This is pure quantum coherence effect since the total
$\bar p p \to \bar\Lambda \Lambda$ cross sections differ by $\sim 30\%$ only at $p_{\rm lab}=2$ GeV/c
(Fig.~\ref{fig:sigma_Lbar_L}) while the hypernuclear production cross sections differ by almost one order of magnitude for Set 1 and Set 2.

\begin{figure}
\begin{center}
   \includegraphics[scale = 0.3]{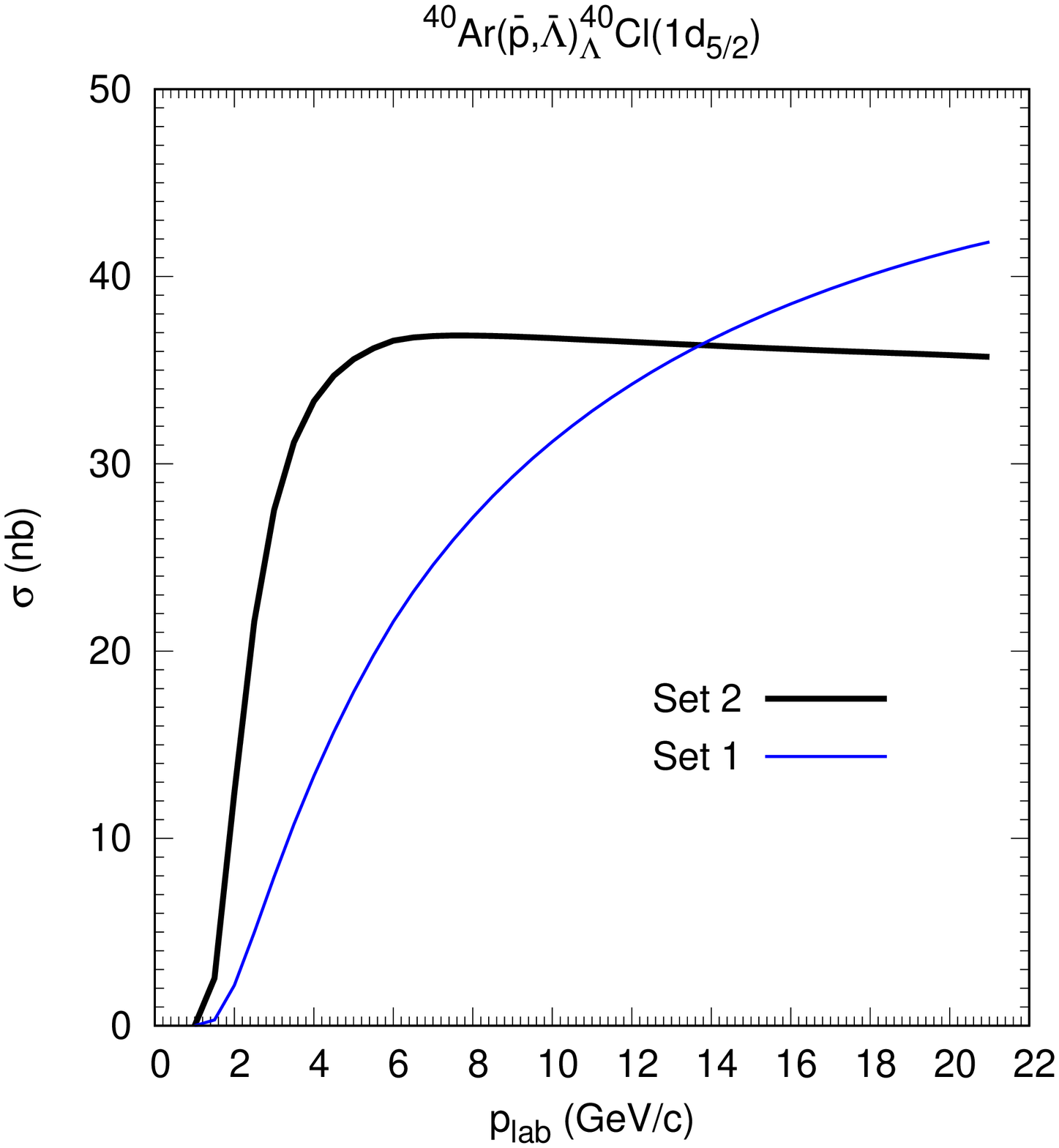}
   \includegraphics[scale = 0.3]{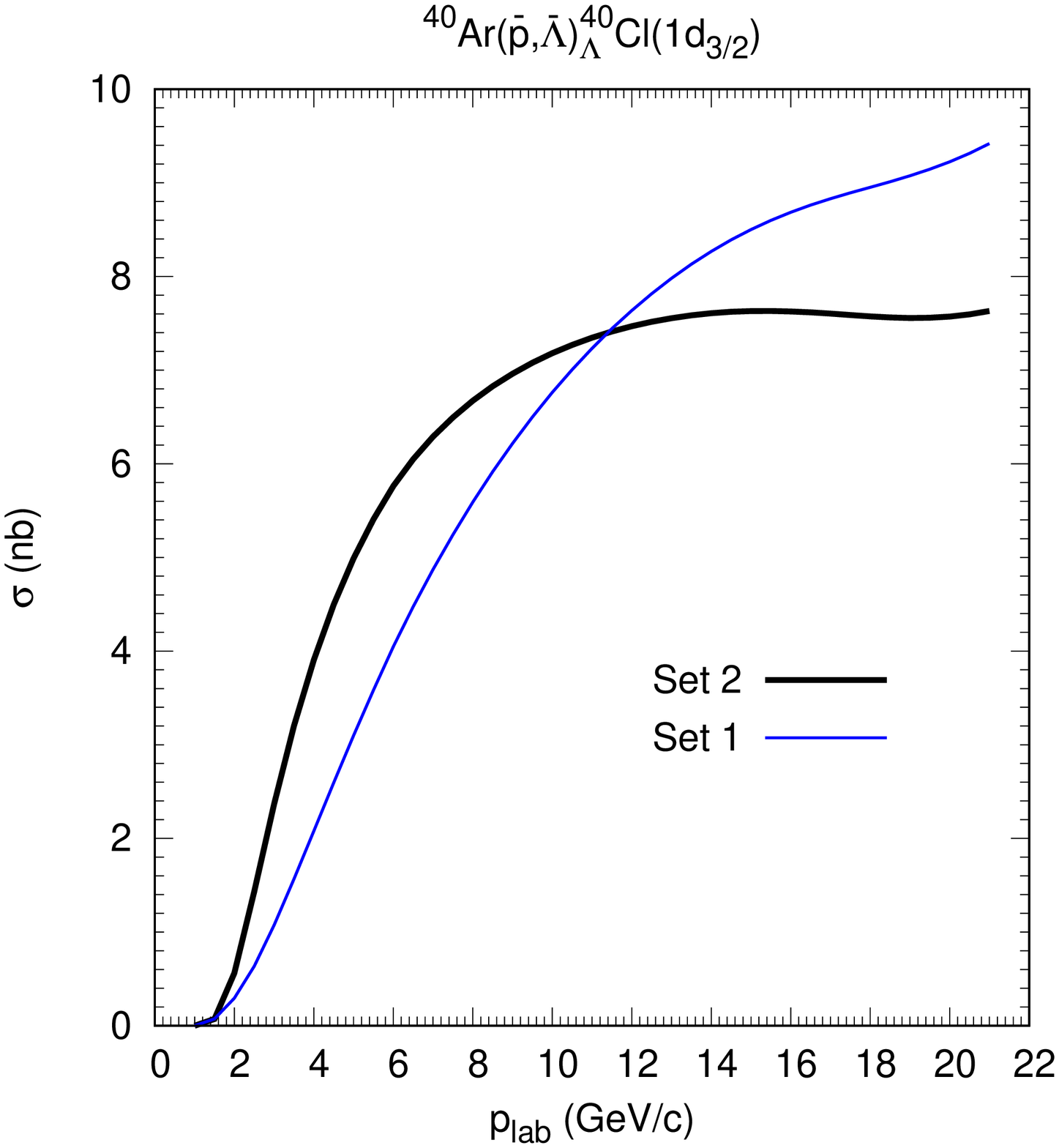}
   \includegraphics[scale = 0.3]{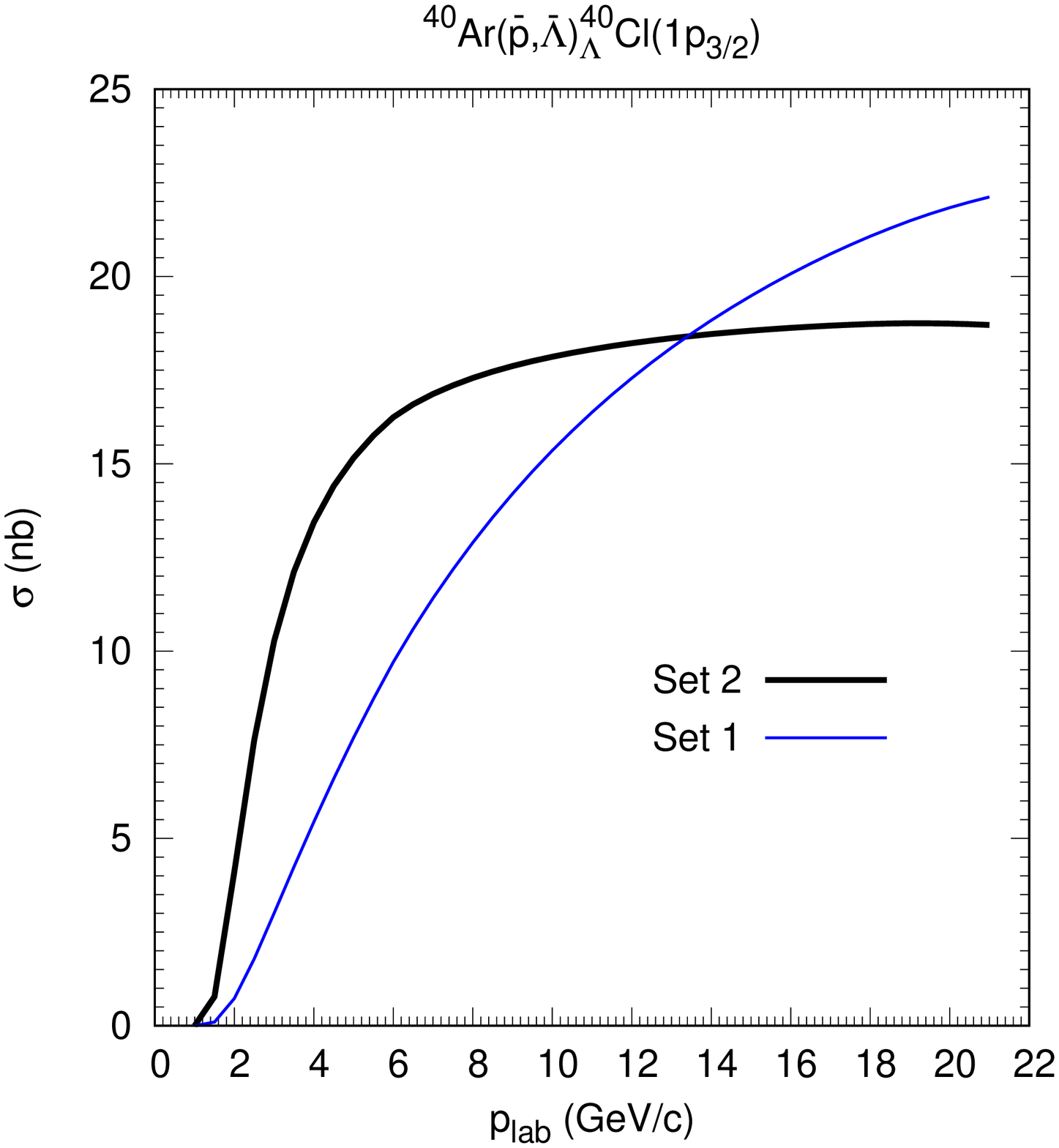}
   \includegraphics[scale = 0.3]{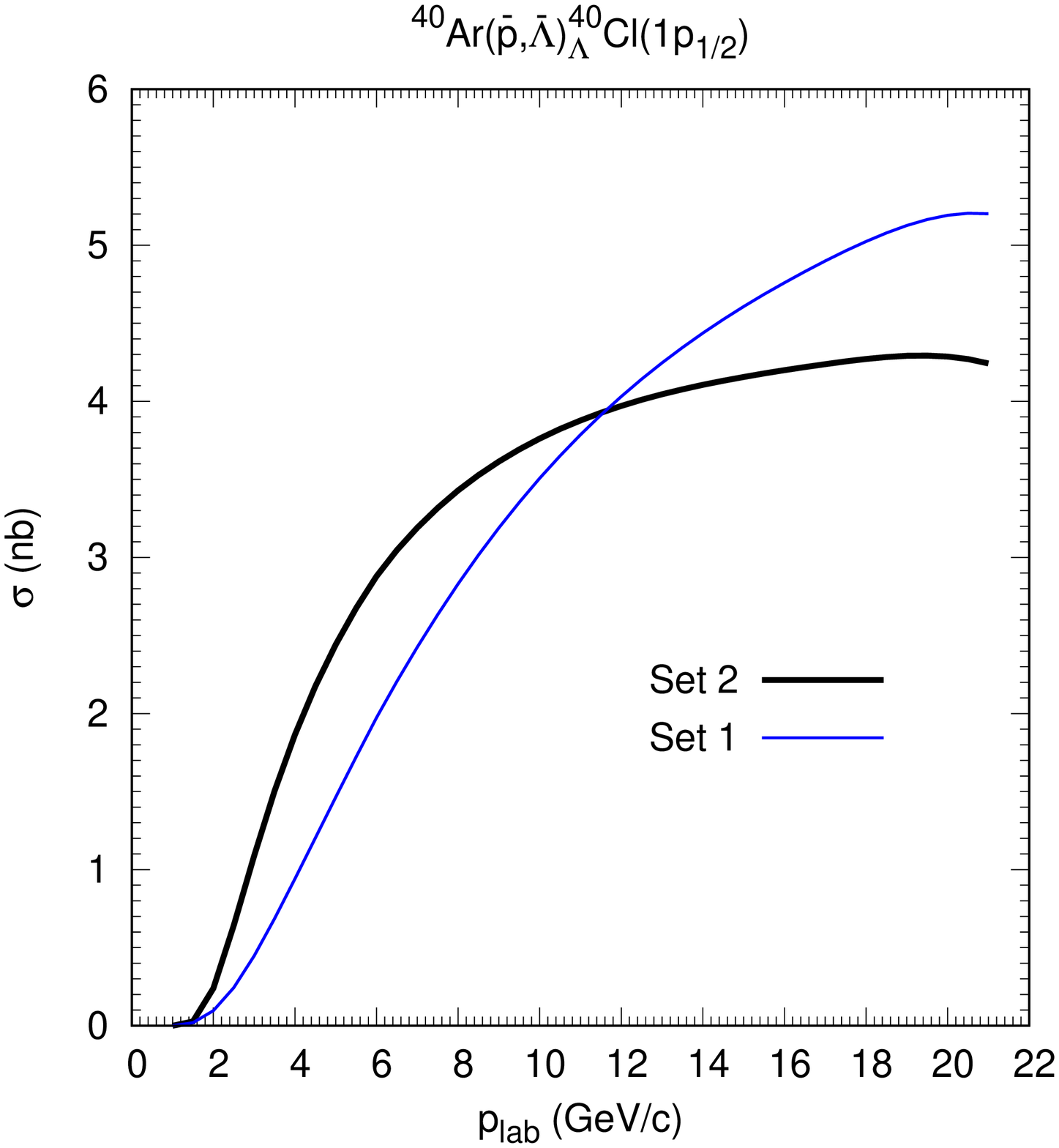}
   \includegraphics[scale = 0.3]{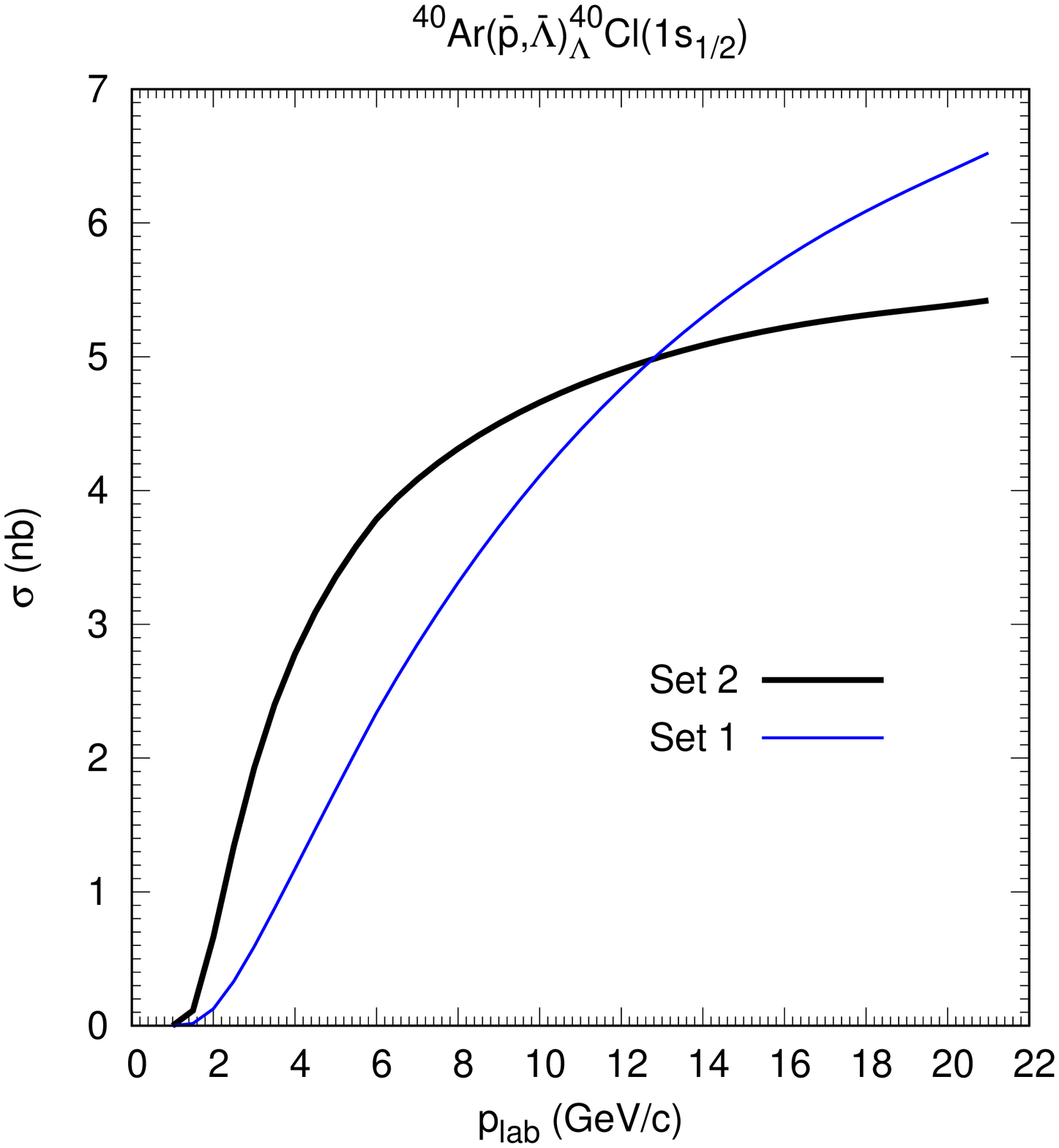}
   \includegraphics[scale = 0.3]{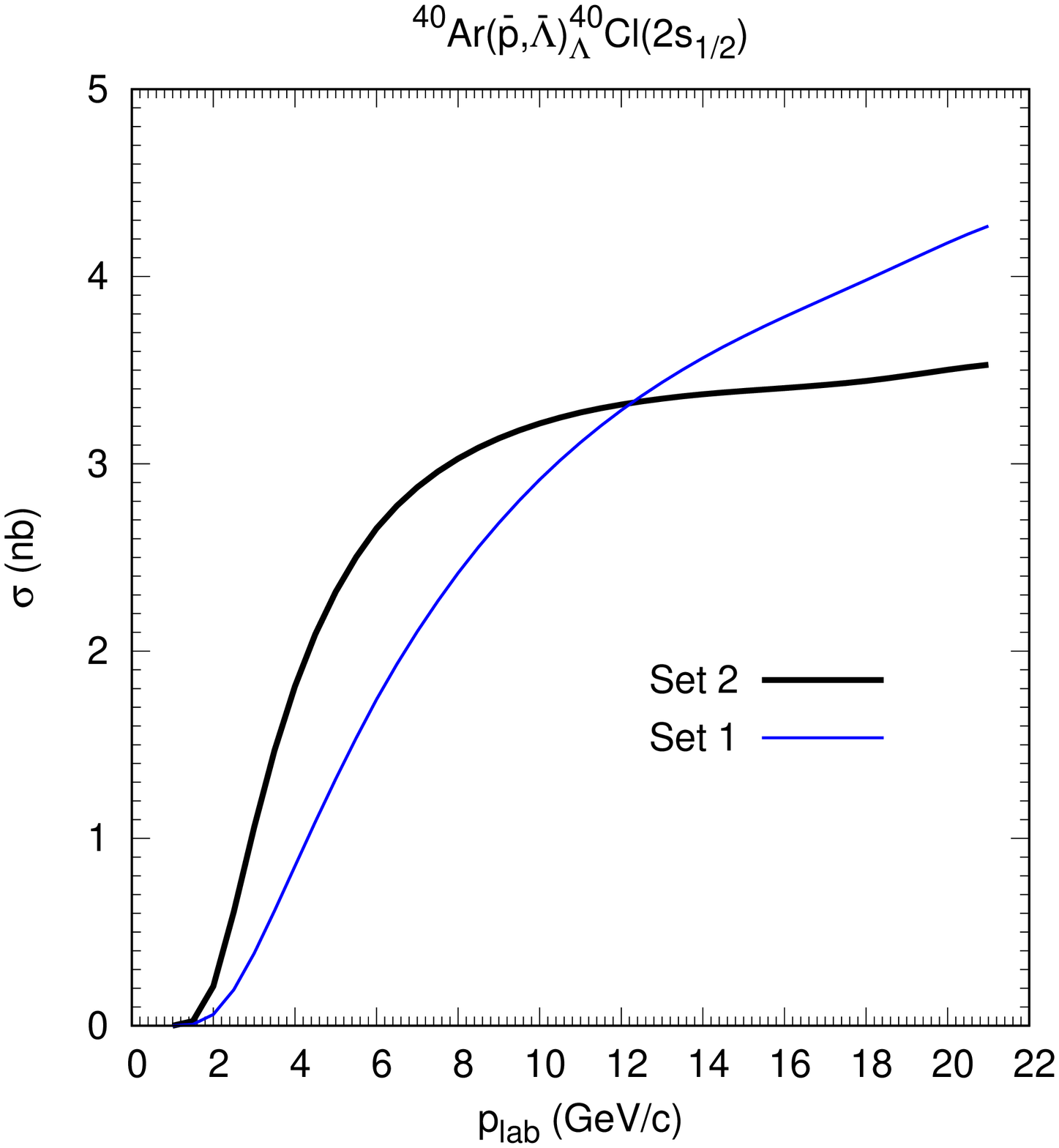}
\end{center}
\caption{\label{fig:md} Beam momentum dependence of the ${}^{40}_{~\Lambda}\mbox{Cl}$ hypernucleus production cross section
  in $\bar{p}\,{}^{40}\mbox{Ar}$ collisions. The thick and thin solid lines show, respectively, the results with (Set 2)
  and without (Set 1) $\kappa$ exchange. Different panels display calculations for different $\Lambda$ states,
  as indicated.}
\end{figure}

The robust signal of the $\kappa$ exchange is visible in the momentum dependence of the hypernucleus
production cross section, as seen in Fig.~\ref{fig:md}.
In calculations without $\kappa$ the cross section is dominated by $K^*$ exchange which leads to a growing cross section with increasing beam energy.
Using set 2, the $\kappa$ meson dominates at moderate beam momenta $\sim 1.5\div 3$ GeV/c (right Fig.~\ref{fig:sigma_Lbar_L}).
Its contributions are seen as a characteristic shoulder in $p_{\rm lab}$-dependence of the hypernuclear production cross section
and even as the appearance of the maximum for $1d_{5/2}$ $\Lambda$ state.

\begin{figure}
\begin{center}
   \includegraphics[scale = 0.3]{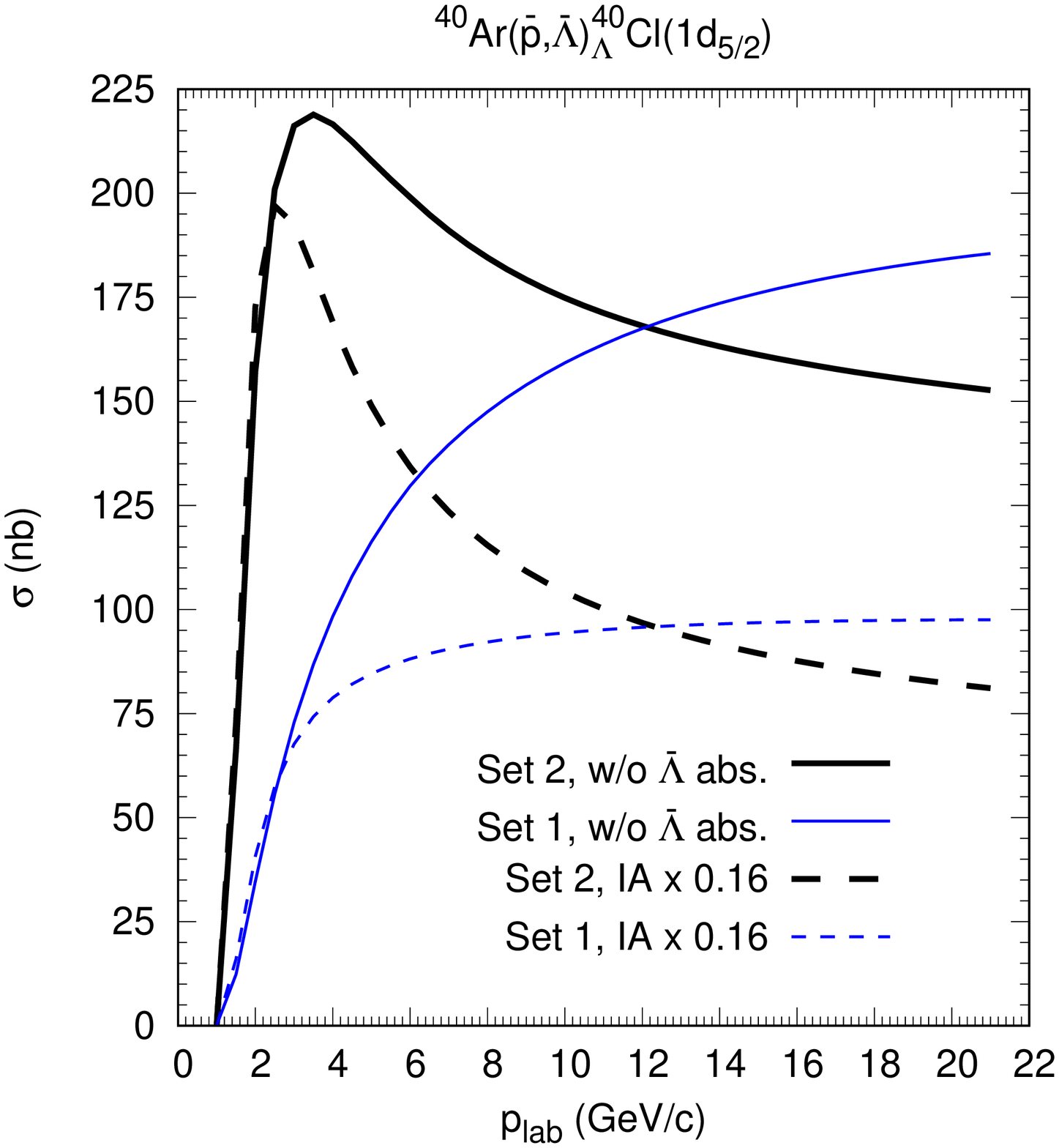}
\end{center}
\caption{\label{fig:tstAbs_md} Same as in Fig.~\ref{fig:md} for $1d_{5/2}$ $\Lambda$ state,
  but for calculations without antihyperon absorption and using IA, as indicated.
  In the case of IA, shown are the cross sections multiplied by 0.16.}
\end{figure}
In Fig.~\ref{fig:tstAbs_md} we present the results for
the beam momentum dependence of the ${}^{40}_{~\Lambda}\mbox{Cl}(1d_{5/2})$ hypernucleus production cross section obtained
by neglecting the absorption of $\bar \Lambda$ and by using the IA. In Set 2 calculation without
$\bar\Lambda$ absorption, the maximum in the $p_{\rm lab}$ dependence  shifts to smaller beam momenta and becomes
more sharp. The same effect is reached by further removing the $\bar p$ absorption (IA calculation). Thus, removing
initial/final state absorption makes the difference between Set 1 and Set 2 calculations even stronger.
Therefore this difference is a clean manifestation of the $\kappa$ exchange and not an artifact of particular approximation
for the ISI/FSI effects.

\section{Summary and conclusions}
\label{SumConcl}

In the present work, the coherent hypernucleus production in $\bar pA$ collisions was investigated.
The production dynamics of the elementary and the in-medium annihilation amplitudes were described in a covariant meson exchange approach.
ISI and FSI of the scattered baryons have been taken into account by eikonal theory.
Baryon bound states were obtained by a variational approach using a RMF-model.
The approach was applied to hypernuclear production in coherent reactions on the medium-heavy ${}^{40}\mbox{Ar}$ target nucleus in the momentum range
$p_{\rm lab}\sim 1.5\div20$~GeV/c. It has been found that the total hypernucleus production cross sections populating a fixed quantum state generally
grow with increasing beam momentum from several nb to a few of 10~nb with a certain sensitivity on the $\Lambda$ bound state. Dynamics of the
${}^{40}\mbox{Ar}(\bar p,\bar\Lambda){}^{40}_{~\Lambda}\mbox{Cl}$ reaction on the $1d_{3/2}$ valence shell proton favors the production of $\Lambda$
states with $j=l+1/2$ with the cross sections increasing with $l$.

We have demonstrated that the pseudoscalar ($K$) exchange is strongly suppressed for the reactions replacing a bound proton by a bound $\Lambda$ hyperon.
Thus, the production mechanism is governed by the exchange of natural parity vector and scalar strange mesons. Including only $K^*$ exchange produces
smooth and structure-less cross sections increasing steadily with beam momentum for all possible bound $\Lambda$ states. However, if the exchange of
the scalar $\kappa$ meson is taken into account we find that at the beam momenta in the range of $p_{\rm lab}=4\div6$~GeV/c a rather sudden transition
from increase to saturation occurs or, as in the case of the $1d_{5/2}$ $\Lambda$ state, a maximum is emerging. These results strongly suggest
that the coherent hypernuclear production in $\bar p A$ annihilation reactions could be a suitable tool to test in quite detail the dynamics
of the production process, down to the possibility to identifying contributions from scalar $\pi K$ correlation as described by the $\kappa$ meson.
As mentioned before, the planned \={P}ANDA@FAIR experiment would be a suitable facility for such studies but experiments could be performed also at J-PARC
if the occasionally discussed antiproton option will be realized.

The theoretical methods sketched above are of general character. With the appropriate choice of parameters
they can be applied to any kind of coherent hyperon production process on nuclei,
in particular, to the process $(\bar p, \Lambda)$ with the capture of $\bar \Lambda$ in the residual nucleus, which requires large momentum
transfer to the struck proton. It is clear that the shell model description of the nuclear ground state is absolutely necessary for the description
of such processes.
However, we would also like to mention that a wide class of hard semiexclusive processes,
such as $(p,pp)$, $(\bar p, \bar p p)$, $(\bar p, \bar \Lambda \Lambda)$, related to the color transparency studies,
might be sensitive to the shell model treatment of the nuclear ground state \cite{Frankfurt:1994nn} and can be studied
theoretically with similar methods.

\begin{acknowledgments}
  This work was supported by the Deutsche Forschungsgemeinschaft (DFG) under Grant No. Le439/9.
  Stimulating discussions with M. Bleicher and M. Strikman are gratefully acknowledged.
\end{acknowledgments}

\bibliography{pbarHyp}

\end{document}